\documentclass[12pt,letterpaper,hyper]{JHEP3}
\usepackage{amsmath,epsfig}
\usepackage{amssymb,amsfonts}
\usepackage{graphicx}
\usepackage{multirow,cite}
\usepackage{subfigure}
\usepackage{axodraw}

\title{New Perspectives on Attractor Flows and Trees from CFT}

\preprint{ASC-LMU 69/10}

\author{
Suresh Nampuri$^{1}$, and Masoud Soroush$^{1,2}$\\

\it $^1$Arnold Sommerfeld Center, Ludwig-Maximilians-Universit\"{a}t\\
\it Theresienstr. 37, M\"{u}nchen 80333, Germany\\

\it $^2$Excellence Cluster Universe, Technische Universit\"{a}t M\"{u}nchen\\
\it Boltzmannstr. 2, Garching 85748, Germany\\

\textrm{Emails}:suresh.nampuri@physik.uni-muenchen.de, masoud.soroush@physik.uni-muenchen.de, %{smurthy@lpthe.jussieu.fr}\\
}

\abstract{In this note, we use  the first order formalism for attractor flows in ${\cal N}=2$ SUGRA extremal black hole backgrounds to establish a formal correspondence between the RG flow of moduli in the underlying ${\cal N}=(2,2)$ SCFT and bulk attractor flows. Starting from a study of  moduli flow trajectories in the CFT, we derive a potential which generates the aforesaid flow. This potential is shown to be a symplectic invariant with the same form as the black hole potential which drives the attractor flows in the bulk. We use these results to make comments on the non-renormalization of extremal black hole entropy and indicate a similar correspondence for two-centered forked flows in CFT.}


\newcommand{\bi}{\begin{itemize}}
\newcommand{\ei}{\end{itemize}}

\renewcommand{\Im}{\mbox{Im}}

\def\h{\eta}

\def\1F1{{}_1\!F_1}
\def\2F0{{}_2\!F_0}

\def\sl{$SL(2,R)$}

\def\h3{$\textrm{H}_3^+$}
\font\manual=manfnt
\def\dbend{\lower3.5pt\hbox{\manual\char127}}

\def\bar{\overline}

\def\rt2{\sqrt{2}}
\def\irt2{{1\over\sqrt{2}}}

\font\cmss=cmss10 \font\cmsss=cmss10 at 7pt
\def\IL{\relax{\rm I\kern-.18em L}}
\def\IH{\relax{\rm I\kern-.18em H}}
\def\rlx{\relax\leavevmode}
\def\ZZ{\rlx\leavevmode\ifmmode\mathchoice{\hbox{\cmss Z\kern-.4em Z}}
 {\hbox{\cmss Z\kern-.4em Z}}{\lower.9pt\hbox{\cmsss Z\kern-.36em Z}}
 {\lower1.2pt\hbox{\cmsss Z\kern-.36em Z}}\else{\cmss Z\kern-.4em
 Z}\fi}
\def\rt2{\sqrt{2}}
\def\irt2{{1\over\sqrt{2}}}

\newcommand{\bea}{\begin{eqnarray}}
\newcommand{\eea}{\end{eqnarray}}
\newcommand{\be}{\begin{equation}}
\newcommand{\ee}{\end{equation}}
\newcommand{\ben}{\begin{eqnarray*}}
\newcommand{\een}{\end{eqnarray*}}
\newcommand{\bem}{\begin{pmatrix}}
\newcommand{\eem}{\end{pmatrix}}
\newcommand{\bl}{\begin{align}}
\newcommand{\el}{\end{align}}

\def\h{\eta}

\begin{document}
\section{Introduction}
Black holes constitute important laboratories for testing the formalisms and predictions for gravity, in
particular, for those emerging from string theory. String theory has made significant advances in the
understanding of various aspects related to extremal black holes. These include an exact microscopic counting of
the entropy of supersymmetric black holes, (See \cite{Dabholkar:2004yr,Dabholkar:2005by,Dabholkar:2005dt,Dabholkar:2008zy,Dabholkar:2009dq,Dijkgraaf:1996it,Mandal:2010cj,Pioline:2005vi,Sen:2005ch,Sen:2008ta,Shih:2005uc,
Strominger:1990pd,Dabholkar:2008tm,Dabholkar:2006xa,Dabholkar:2007vk}.) and in recent times, an improved understanding of the
entropy of non-supersymmetric extremal black holes (See \cite{Guica:2008mu,Jejjala:2009if,Nampuri:2007gw}.). A related phenomenon in  case of extremal black
holes that has been the focus of extensive research in the past decade is the attractor mechanism.
 Given an extremal black hole solution of a supergravity action with scalar moduli, the attractor mechanism fixes
 the moduli that are non-constant in this background at the horizon in terms of the charges of the black holes.
 Hence their horizon values (referred to as attractor values) are independent of their asymptotic values. This
 mechanism was discovered and explored for supersymmetric (SUSY) black holes in \cite{Ferrara:1995ih,Ferrara:1996dd,Ferrara:1996um} and later, for non-supersymmetric
 (non-SUSY) black holes, firstly in \cite{Ferrara:1997tw}, and followed by \cite{Goldstein:2005hq,Tripathy:2005qp,Kallosh:2005ax,Kallosh:2006bt,Kallosh:2006bx,Sen:2007qy,Goldstein:2009cv,Goldstein:2010aw}.
 The connection between the attractor mechanism and the entropy can be summarized elegantly via the entropy
 function formalism (See \cite{Sen:2007qy} and references therein.). The entropy function is essentially the Hamiltonian obtained from the 1D effective
 Lagrangian density, which is derived from the 4D action by substituting the near-horizon solution of
 the black hole into the action. The entropy function is therefore a functional of the moduli and the charges of
 the black hole. Extremizing the entropy function w.r.t the various moduli give the extremum or
 attractor values of the moduli at the horizon in terms of the charges while the moduli that are flat directions
 are unaffected by this extremization process. The extremized values of the moduli when substituted back into the
 entropy give its extremum value which is the entropy of the black hole. Hence the horizon values of the moduli
 act as attractors in moduli space. Physically, the attractor mechanism can be explained in terms of the near-horizon geometry of the extremal black hole which always has an $AdS_2$ part. This results in an infinite radial throat which allows all fluctuations in the asymptotic value of the moduli to die down as it travels down the infinite throat so that the horizon value is unaffected by deviations in the asymptotic value. This is a generic feature of extremality and presupposes no supersymmetry. This strongly indicates that there must be a universal formalism for extremal attractors in SUGRA, independent of supersymmetry arguments, and this is the main avenue of discussion that forms the motivation and theme of this paper.
 \par
 The attractor mechanism imposes a boundary condition on the values of the moduli at the horizon. This implies
that the moduli can obey first order equations of motion. These equations have been derived for supersymmetric
black holes in (\cite{Denef:2000nb,Ferrara:1997tw})\footnote{For more on supersymmetric attractors, see \cite{Moore:1998pn,Moore:1998zu}.}. In this case, the radial derivative of the moduli is proportional to the derivative of
the absolute value of the central charge $|Z|$ w.r.t. the moduli. Both quantities vanish at the horizon where $|Z|$
gets
 extremized, and the square of its extremum value gives the entropy. Exact solutions for the attractor flow of moduli in ${\cal N}=2$ SUGRA can be found in \cite{Bates:2003vx} and \cite{Behrndt:1997ny}.
 \par
 In this note, we use a proceedure developed in \cite{Ceresole:2007wx,Bellucci:2008sv,Ceresole:2009vp,Andrianopoli:2010bj} to write down the first order equations of motion for the moduli in the non-supersymmetric case.
  Their functional form is exactly the same as in the supersymmetric case with $|Z|$ replaced by a
 positive definite function-$|R|$.\footnote{This function was referred to as the fake superpotential W in \cite{Andrianopoli:2007gt} who discuss  aspects of the first order equations. The first order Bogmol'yni trick for obtaining the equations and the fake superpotential in the non-SUSY case was developed in \cite{Ceresole:2007wx}  and later in \cite{Cardoso:2007ky} and \cite{Bellucci:2008sv}. Here we give an explicit construction of the function for completeness.} $|R|$ gets minimized at the horizon and its minimum value gives the
 entropy. Hence, just as the attractor values of the moduli in a SUSY black hole background can be regarded as
 solutions of $\partial |Z| = 0$\footnote{Here the partial derivative is w.r.t. the moduli.}, the attractor values
 in a non-SUSY background can be regarded as solutions of $\partial |R| = 0$.  $|R|$ can be constructed
 from $|Z|$ by operating on the charges in such a way as to flip the sign of the duality invariant constructed
 from the charges. Hence, by placing the attractor formalisms for the SUSY and non-SUSY backgrounds on the same
 footing, we can, given a SUSY attractor value , perform the same operation on the charges in it as required to go
 from $|Z|$ to $|R|$ and the resulting value will be a non-SUSY attractor value. Further, given a technique of
 obtaining the general attractor flow from the attractor values (replacing charges with harmonics) in the SUSY
 case, the same technique must apply for non-SUSY cases since the functional forms of the equations are the same.
 \par
 We use the first order attractor formalism to develop and explore the idea of split attractor flows in the non-SUSY case and find that the remarkable similarity between the SUSY and non-SUSY formalisms allows us to
 use technology available for the analysis of SUSY split attractor flows in SUGRA ala \cite{Bates:2003vx} and \cite{Denef:2000nb,Galli:2009bj} for the non-SUSY case. We end the SUGRA section of this note with comments and discussion on this formalism.
 \par
 We then proceed in the subsequent sections to draw a correspondence between the moduli flow equations in CFT
 (derived in \cite{Brunner:2010xm}) and the supergravity attractor equations. This exact correspondence between
 supergravity flows and moduli flows in strongly coupled gravity described by the underlying CFT is then used to
 indicate statements about the non-renormalization of the entropy in the non-supersymmetric case. A 'physical'
 explanation of this phenomenon is given in terms of the 4D/5D lift in the summary section.\footnote{For details on attractors in higher ${\cal N}$ and higher dimensional SUGRA refer to \cite{Ceresole:2010hq,Ceresole:2009id,Andrianopoli:2008ea,Ferrara:2010ru}.}

\section{Background: Black Hole Attractors}
The bosonic part of the action of the four dimensional ${\cal N}=2$ supergravity consisting of $N+1$ abelian
$U(1)$ vector multiplets and no hypermultiplet is given by
\begin{eqnarray}\label{action}
S=\frac{1}{16\pi}\int d^{4}x\sqrt{|G|}\Big(-\frac{{\mathcal{R}}}{2}+g_{a\bar{b}}\,\partial
z^{a}\cdot\partial\bar{z}^{\bar{b}}+ \mbox{Im}({\mathcal{N}}_{AB}){\mathcal{F}}^{A}\cdot{\mathcal{F}}^{B}+
\mbox{Re}({\mathcal{N}}_{AB}){\mathcal{F}}^{A}\cdot(*{\mathcal{F}})^{B}\Big)\ ,
\end{eqnarray}
where ${\cal F}^{A}$ is the field strength associated with the gauge field ${\cal A}^{A}$, and
${\mathcal{N}}_{AB}$ is the gauge kinetic function. The fermionic part of the action can be constructed via
supersymmetry transformations. The structure of the vector multiplet moduli space of the theory, which is a
special K\"{a}hler manifold, is governed by $N=2$ special geometry \cite{Strominger:1990pd}. The basic ingredient
for describing the moduli space of the classical ${\cal N}=2$ supergravity theory is the prepotential $F(X)$ which
is a holomorphic homogeneous function of degree two and is given by
\begin{eqnarray}\label{prepot}
F(X)=d_{ABC}\,\frac{X^{A}X^{B}X^{C}}{X^{0}}\ ,
\end{eqnarray}
where $X^{A}$'s are the homogeneous coordinates of the moduli space and $d_{ABC}$ is a completely symmetric
tensor. The positive definite metric $g_{a\bar{b}}$ on the moduli space is then derived from the K\"{a}hler
potential. For the classical prepotential (\ref{prepot}), the K\"{a}hler potential turns out to be
\begin{eqnarray}\label{Kahlerpot}
K=-\log\Big(i\,d_{ABC}(X^{A}-\bar{X}^{A})(X^{B}-\bar{X}^{B})(X^{C}-\bar{X}^{C})\Big)\ .
\end{eqnarray}
The homogeneous coordinates of the moduli space, $X^{A}$, are related to the inhomogeneous coordinates
$z^{a}=\frac{X^{a}}{X^{0}}$ via a choice of the gauge. In this note, we will work with the gauge in which $X^{0}$
is real. It is also convenient to define a covariant derivative on the moduli space. The covariant derivative
$D_{a}$ acts on a given field $\Psi$ as
\begin{eqnarray}\label{Covder}
D_{a}\Psi=\left(\partial_{a}+\frac{p}{2}\,\partial_{a}K\right)\Psi\ ,
\end{eqnarray}
where $p$ is the chiral $U(1)$-weight of $\Psi$, and it is determined by the K\"{a}hler transformation property of
$\Psi$.
\par
The most general stationary spherically symmetric single center extremal black hole solution of ${\cal N}=2$
supergravity is obtained by considering the following ansatz for the spacetime metric
\begin{eqnarray}\label{metric}
ds^2=-e^{2U(r)}dt^{2}+e^{-2U(r)}dx^idx^i\ .
\end{eqnarray}
In this background, it is then straightforward to define the symplectic set of electric and magnetic charges
$(q_{A},p^{A})$ of the single center black hole configuration. This gives rise to the following superpotential for
the theory
\begin{eqnarray}\label{superpot}
W=q_{A}X^{A}-p^{A}\partial_{A}F\ .
\end{eqnarray}
In the context of type II string compactifications on Calabi-Yau manifolds, ${\cal N}=2$ supergravity arises as
the low energy effective theory in four dimensions. In particular, in type IIA theory, the electric and magnetic
charges $(q_{0},q_{a},p^{a},p^{0})$ correspond to the charges of $(D0,D2,D4,D6)$-branes wrapping
$(0,2,4,6)$-cycles of the compact Calabi-Yau manifold. Moreover, the completely symmetric tensor $d_{ABC}$ defined
in (\ref{prepot}) corresponds to the intersection number of the two-cycles of the internal  Calabi-Yau space.
\par
The central charge of the supersymmetry algebra is specified in terms of the superpotential $W$ and the K\"{a}hler
potential $K$, and is given by $Z=e^{K/2}\,W$. Plugging the ansatz (\ref{metric}) into the bosonic action
(\ref{action}) yields the following effective lagrangian (for details see \cite{Ferrara:1997tw} as well as
appendix A of \cite{Kallosh:2006bt}) for the single center black hole
\begin{eqnarray}\label{efflag}
{\mathcal{L}}(U(\tau),z^{a}(\tau),\bar{z}^{\bar{a}}(\tau))=
\dot{U}^{2}+g_{a\bar{b}}\,\dot{z}^{a}\dot{\bar{z}}^{\bar{b}}+e^{2U}V_{BH}\ ,
\end{eqnarray}
where $\tau$ denotes the inverse of the radial coordinate $r$ ($\tau=\frac{1}{|r-r_h|}$) and the dot denotes the
differentiation with respect to $\tau$. Furthermore, $V_{BH}$ in (\ref{efflag}) is the black hole potential and is
defined in the following way
\begin{eqnarray}\label{BHpot}
V_{BH}=|DZ|^{2}+|Z|^{2}\equiv g^{a\bar{b}}(D_{a}Z)\bar{D}_{\bar{b}}\bar{Z}+Z\bar{Z}=4g^{a\bar{b}}\partial_{a}|Z|
\partial_{\bar{b}}|Z|+|Z|^{2}\
.
\end{eqnarray}
\footnote{In (\ref{BHpot}), we  use the identity- $D_{a}Z=2\,e^{i\alpha}\partial_{a}|Z|$ (with $\alpha$ being the
phase of the central charge) to express the second term of the potential in terms of partial derivatives.} The
equations of motion resulting from the one-dimensional action give the metric field and the moduli as functions of
the radial coordinate. Now, for extremal blackholes, these solutions exhibit a remarkable property. Essentially,
the solutions for the metric field $e^{-U}$ and the moduli end at values on the horizon which are purely fixed in
terms of the charges of the black hole and which are completely independent of the asymptotic boundary conditions
imposed on the metric and moduli. The horizon values of the moduli are fixed purely in terms of the electric and
magnetic charges by minimizing the above effective potential (\ref{BHpot}). Hence, the flow in moduli
space from asymptotic infinity terminates at a fixed point, represented by the horizon values-the black hole
serves as an attractor point in moduli space and this property is referred to as the attractor mechanism, while
the near-horizon $AdS_2$ geometry for extremal black holes where the moduli and the metric field get fixed at
their horizon values is referred to as the attractor geometry.
 We now proceed to a systematic study of this mechanism for both SUSY and non-SUSY extremal black holes.
\section{Attractor Equations}
Now, given a dynamical system satisfying a second order differential equation, we require two input conditions to
solve the dynamical equation-say, the initial value problem specified by the value of the dynamical quantity at a
point and its derivative at that point, or a boundary value problem with the value of the quantity specified at
two points in its configuration space. For the case of the moduli in our theory of interest, we could take these
two points to be at asymptotic infinity and at the horizon. However, since the derivatives of the moduli w.r.t. the
radial coordinate vanish at the attractor and the attractor values are fixed for a given charge configuration, we
need to specify only one initial condition for the equation to be solved and hence we expect the dynamics of the
1D Lagrangian to be captured by first order equations of motion for the metric field and the moduli. We first
write down the second order equations arising from the Lagrangian.
\begin{eqnarray}
   &&\ddot{U}=e^{2U}V_{BH}\ ,\label{attract1}\\
   &&\ddot{z}^{a}+\Gamma^{a}_{bc}\dot{z}^{b}\dot{z}^{c}=e^{2U}g^{a\bar{b}}
   \partial_{\bar{b}}V_{BH}\ .\label{attract2}
\end{eqnarray}
There is a constraint equation arising from the zero Hamiltonian constraint at extremality
\begin{equation}\label{constr}
   \dot{U}^{2}+g_{a\bar{b}}\,\dot{z}^{a}\dot{\bar{z}}^{\bar{b}}-e^{2U}V_{BH}=0\ .
\end{equation}
Now, the attractor mechanism implies that there exists scalar functions of the moduli,
$R_{1}(z^a,\bar{z}^{\bar{a}})$ and $R_{2}(z^a,\bar{z}^{\bar{a}})$, satisfying
\begin{eqnarray}
   &&\dot{U}\propto R_{1}(z,\bar{z})\ ,\\
   &&\dot{z^a}\propto g^{a\bar{b}}\partial_{\bar{b}}R_{2}(z,\bar{z})\ ,
\end{eqnarray}
with $\dot{z^a}=0$ at the horizon.\footnote{Note that $R_1$ must be a real function since U is real, and that both
$R_1$ and $R_2$ are functions of both the holomorphic moduli $z^a$ and their complex conjugates.} Let us therefore
take the first order equations to be formally
\begin{eqnarray}
   \dot{U}&=& -e^{U}R_{1}(z,\bar{z})\ ,\label{foatt1}\\
   \dot{z^a}&=&-2 e^{U}g^{a\bar{b}}\partial_{\bar{b}}R_{2}(z,\bar{z})\ .  \label{foatt2}
\end{eqnarray}
Differentiating the first equation, (\ref{foatt1}), and comparing with (\ref{attract1}), we find that
\begin{equation}
V_{BH}=(R_{1}(z,\bar{z}))^2-e^{-U}\dot{R_1}(z,\bar{z})\ .
\end{equation}
Substituting $\ddot{U}=e^{2U}V_{BH}$ in (\ref{constr}) and using the expression for the potential in terms of
$R_{1}$, we have,
\begin{equation}
\ddot{U}-\dot{U}^2=g_{a\bar{b}}\,\dot{z}^{a}\dot{\bar{z}}^{\bar{b}}=-e^{U}
\dot{R}_{1}(z,\bar{z})=-e^{U}\left(\dot{z}^{a}\partial_{a}R_{1}(z,\bar{z})+\dot{\bar{z}}
^{\bar{a}}\partial_{\bar{a}}R_{1}(z,\bar{z})\right)\ .
\end{equation}
Comparing the above equation with (\ref{foatt2}), we realize that $R_{1}=R_{2}$. Therefore, we drop the subscripts
and denote the real scalar function that appears in the first order equations as $|R(z,\bar{z})|$. The final first
order attractor equations are therefore-
\begin{eqnarray}
   \dot{U}&=& -e^{U}|R(z,\bar{z})|\ ,\label{firstU}\\
   \dot{z}^a&=&-2\,e^{U}g^{a\bar{b}}\partial_{\bar{b}}|R(z,\bar{z})|\ ,\label{firstz}
\end{eqnarray}
Where $R$ enters the potential
  \footnote{Here we use the first order equations and
  $\dot{|R|}=\dot{z}^a\partial_{a}|R|+\dot{\bar{z}}^{\bar{a}}\partial_{\bar{a}}|R|$  to substitute for $\dot{|R|}$ in
  $V_{BH}$.}
  as
   \begin{equation}\label{potR}
   V_{BH}=4\,g^{a\bar{b}}\partial_{a}|R|\,\partial_{\bar{b}}|R|+|R|^{2}\ .
   \end{equation}
Now using the first order equations of motion, (\ref{firstU}) and (\ref{firstz}), we find that the second order
differential equation for moduli, (\ref{attract2}), is automatically satisfied\footnote{\ Since the vector
multiplet moduli space is a K\"{a}hler manifold, one has the identity
$\partial_{\bar{c}}g^{a\bar{b}}=g^{a\bar{d}}g_{\bar{c}e}\partial_{\bar{d}}g^{e\bar{b}}$  for the K\"{a}hler
metric. One needs to use this identity to prove that (\ref{attract2}) is fulfilled.}. It is also worth noting that
once we have the potential (\ref{potR}), we can perform the usual Bogomolyni trick \cite{Ceresole:2007wx} of completing squares
\begin{eqnarray}\label{Bogomol}
S_{1D}=\int_{0}^{\infty} d\tau\Big((\dot{U}\pm
e^{U}|R|)^{2}+\big|\dot{z}^{a}\pm2e^{U}g^{a\bar{b}}\,\partial_{\bar{b}}|R|\big|^{2}\Big)
\pm2e^{U}|R|
\Big|_{\tau=0}^{\tau=\infty}-2(e^{U}|R|)\Big|_{\tau=\infty}\ .
\end{eqnarray}
In the supersymmetric case, the $R$-function is nothing except $|Z|$ and one reproduces the well-known 1D
effective action derived in \cite{Denef:2000nb} and \cite{Ferrara:1997tw}. For non-SUSY black holes, the form of
the effective action and hence the first order equations of motion remain the same, but the $R$-function for the
non-SUSY case is not equal to the absolute value of the central charge. In the next subsections, we explicitly
construct the $R$-function for non-SUSY cases.

\subsection{First order SUSY attractor equations}
Now, for the SUSY blackholes, as is well known, $|Z|$ is minimized during the attractor flow from the boundary to
the horizon and hence $\partial_a|Z|=0$ at the horizon. Hence each of the two terms in the potential is minimized
separately at the horizon and the minimum value of the potential which is also the minimum value of $|Z|^2$ is
equal to the entropy and the square of the Bertotti-Robinson mass (BR). Hence, putting $|R|=|Z|$ in (\ref{Bogomol}) yields
the first order equations for SUSY black holes
\begin{eqnarray}
\dot{U}&=&-e^{U}|Z|\ ,\label{BPSeq1}\\
\dot{z}^{a}&=&-2 e^{U}g^{a\bar{b}}\partial_{\bar{b}}|Z|\ .\label{BPSeq2}
\end{eqnarray}
 For supersymmetric solutions, these are nothing but the first order BPS equations of motion
 \cite{Ferrara:1997tw,Denef:2000nb} which can be obtained by using the usual Bogomol'nyi trick on the Lagrangian.

One clever way of finding the solutions to (\ref{BPSeq1}) and (\ref{BPSeq2}) is to take the advantage of the
horizon values of the moduli fields. Let us define
$\Sigma(\mathbf{\Gamma})=|Z(\mathbf{\Gamma})|\Big|_{z=z_{*}(\mathbf{\Gamma})}$, where
$\mathbf{\Gamma}=(p^{A},q_{A})$ and $z_{*}(\mathbf{\Gamma})$ denotes the horizon values of the moduli. Physically
speaking, $\Sigma(\mathbf{\Gamma})$ is nothing but the entropy of the supersymmetric single center black hole. It
was shown in \cite{Bates:2003vx} that the exact solution for the moduli fields of a supersymmetric single center
black hole can be obtained from the following formula
\begin{eqnarray}\label{Sol}
z^{a}(\tau)=\frac{H^{a}-i\frac{\partial \Sigma(H)}{\partial H_{a}}}{H^{0}-i\frac{\partial \Sigma(H)}{\partial
H_{0}}}\ ,
\end{eqnarray}
in which one substitutes the electric and magnetic charges $\mathbf{\Gamma}$ in $\Sigma(\mathbf{\Gamma})$ by
appropriate harmonic functions. The harmonic functions are defined in terms of $\tau$, the inverse of the radial
coordinate and they are explicitly given by
\begin{equation}\label{Harmfun}
\mathbf{H}(\tau)=\left(H^{A},H_{A}\right)=\left(h^{A}, h_{A}\right)+\left(\Gamma^{A},\Gamma_{A}\right)\tau\ ,
\end{equation}
where $(h^{A},h_{A})$ are the asymptotic values at infinity and are subject to the constraint
$\langle\mathbf{h},\mathbf{\Gamma}\rangle=0$. The solution (\ref{Sol}) together with the solution for the metric
$e^{-2U}=\Sigma(\mathbf{H})$ fulfill the BPS equations (\ref{BPSeq1}) and (\ref{BPSeq2}). The solution for the
vector fields can be constructed from the harmonic functions correspondingly. These solutions all satisfy
$\partial_a|Z|=0$ at the horizon. In the appendix we use this equation to explicitly derive the attractor values of
the moduli for various charge configurations in the STU model. Notice that for the SUSY black hole, the BPS bound
is saturated and $M=|Z|$. Hence one could think of the attractor flow as extremizing the mass function which
assumes the ADM mass value at asymptotic infinity and the B-R mass at the horizon, write the potential as
$V_{BH}=4g^{a\bar{b}}\partial_{a}M\partial_{\bar{b}}M+M^{2}$ and obtain the attractor values as solutions of
$\partial_a M=0$ at the horizon.

\subsection{First order non-SUSY attractor equations}
For extremal non-SUSY black holes, we consider two classes of black holes- those which can be constructed from the
same charge species as SUSY blackholes by flipping the signs of some of the charges-example include the D0-D4,
D0-D2-D4, D2-D6, D0-D4-D6 etc. systems and those which have no such SUSY equivalent-a classic example being the
D0-D6 system.
\par
In order to find the $R$-function for the non-SUSY case, we restrict ourselves initially, for simplicity to the
${\cal N}=2$ STU example and consider configurations with axion-free solutions. In particular, we consider D0-D4
and D2-D6 systems. The full attractor flow for the moduli and the metric field in these cases ca be explicitly
obtained from \cite{Kallosh:2006ib}. We plug the solution for the metric field so obtain (\ref{firstU}). This
gives rise to an explicit form for the $R$-function and a quick observation reveals that the $R$-function is
precisely the absolute value of the central charge, $|Z|$, of the equivalent SUSY charge configuration. This SUSY
configuration can be obtained by reversing the sign of appropriate charges in the original non-SUSY configuration.
In the next subsection, we proceed to explicitly illustrate and consolidate this observation for general charge
configurations in the STU model. The STU model is a four dimensional ${\cal N}=2$ theory with 3 vector multiplets
and no hypermultiplets. Its  Lagrangian can be explicitly constructed from the superpotential (\ref{prepot}) with
$a$,$b$ and $c$ running from 1 to 3 and $d_{abc}=-\frac{1}{6}$. The electric and magnetic charges can be organized into
$SO(2,2)$ lattices
\begin{equation}
(Q|P)=(q_0, -p^1, q_2, q_3|q_1, p^0, p^3, p^2)\ .
\end{equation}
with the norm in charge space given by
\begin{eqnarray}
\frac{Q^2}{2}&=&(-q_0 p^1 + q_2 q_3)\ ,\\
\frac{P^2}{2}&=&(q_1 p^0+ q_2 q_3)\ ,\\
Q\cdot P&=& q_0 p^0 - q_1 p^1 + q_1 p^1 + q_2 p^2\ .
\end{eqnarray}
For future expedience, we define the matrix
\begin{equation}
D_{ab}=D_{abc}p^c=-\frac{1}{6}\left(
                     \begin{array}{ccc}
                       0 & p^3 & p^2 \\
                       p^3& 0 & p^1 \\
                       p^2 & p^1 & 0 \\
                     \end{array}
                   \right)\ .
\end{equation}
The K\"{a}hler potential and superpotential are obtained from (\ref{Kahlerpot}) and (\ref{superpot}) respectively
as
\begin{eqnarray}
K&=&-\log\big(i d_{abc} (z^1-\bar{z}^1)(z^2-\bar{z}^2)(z^3-\bar{z}^3)\big)\ ,\\
W&=&q_0 + q_a z^a + \frac{p^a}{z_a}{z^1 z^2 z^3} - p^0 z^1 z^2 z^3\ .\label{STUsup}
\end{eqnarray}
The metric and the connection on the K\"{a}hler moduli space determined by the three scalars, ${z^a}$
are\footnote{The index $a$ when repeatedly present in a term is summed over while the index $I$ is always a free
variable and never summed over.}
\begin{eqnarray}
 g_{a\bar{b}}&=& - \frac{1}{(z^a-\bar{z}^a)^2}\delta_{ab}\ ,\\
 \Gamma^I_{II}&=&-\frac{2}{z^I-\bar{z}^I}\ .
 \end{eqnarray}
The U-duality symmetry group  operating on the charges and moduli is $(SL(2,\mathbb{R}))^3$. The entropy of the
charge configuration in supergravity is proportional to the square-root of the U-duality invariant ${\cal I}=Q^2
P^2-(Q\cdot P)^2$.\footnote{This is the Caley hyper-determinant formed from the charges in the STU model. See
\cite{Kallosh:2006zs} for details.} SUSY configurations correspond to charges with a positive $\cal{I}$ while
non-SUSY configurations correspond to a negative $\cal{I}$. In general, given a non-SUSY charge configuration, one
can construct the equivalent SUSY configuration (if it exists) by an appropriate change of sign or absolute value
of the charges so as to flip the sign of $\cal{I}$ while keeping its absolute value unchanged. This operation on
the charges is unique modulo the U-duality group. In the ensuing subsections, we consider various special cases of
the most general charge configuration by putting various charges to zero and derive the $R$-function for these
cases. We will restrict ourselves in this note to looking at the direct construction of the $R$-function for axion-free charge systems. General constructions can be seen in \cite{Ceresole:2007wx,Cardoso:2007ky,Gimon:2007mh}.

\subsubsection{D2-D6 system}
As the first explicit example, we proceed to analyze the special case of D2-D6 system. Then, from the general
formula (\ref{STUsup}), the superpotential reads
\begin{eqnarray}\label{W26}
W(z^{1},z^{2},z^{3})=q_{a}z^{a}-p^{0}z^{1}z^{2}z^{3}\ .
\end{eqnarray}
This system was previously analyzed in \cite{Kallosh:2006ib} and both SUSY and non-SUSY solutions have been
constructed. The horizon values of the moduli in the non-SUSY branch are obtained by minimizing $V_{BH}$
\begin{eqnarray}\label{attsol26}
z^{1}=\pm i\sqrt{-\frac{q_{2}q_{3}}{p^{0}q_{1}}}\ ,\ z^{2}=\pm i\sqrt{-\frac{q_{1}q_{3}}{p^{0}q_{2}}}\ ,\
z^{3}=\pm i\sqrt{-\frac{q_{1}q_{2}}{p^{0}q_{3}}}\ .
\end{eqnarray}
The signs of the solutions in (\ref{attsol26}) are taken such that the K\"{a}hler volume
$e^{K}=\frac{i}{8\,z^{1}z^{2}z^{3}}$ is positive.

\par
Now, in order to construct the $R$-function for D2-D6 system, it suffices to flip the sign of the D6-brane charge:
$p^{0}\rightarrow-p^{0}$. We notice that under this sign flip the duality invariant ${\cal
I}=\sqrt{4\,p^{0}q_{1}q_{2}q_{3}}$ is transformed to ${\cal I}=\sqrt{-4\,p^{0}q_{1}q_{2}q_{3}}$. The $R$-function
for non-SUSY solutions of the D2-D6 system is then found
\begin{eqnarray}\label{R26}
|R(z,\bar{z})|=\big|e^{K/2}(q_{a}z^{a}+p^{0}z^{1}z^{2}z^{3})\big|\ .
\end{eqnarray}
First, we notice that by minimizing the $R$-function (\ref{R26}) with respect to moduli, $\partial_{a}R=0$ and
$\partial_{\bar{a}}R=0$, we exactly recover the horizon values (\ref{attsol26}). Second, we proceed to prove that
the $R$-function (\ref{R26}) fulfills the first order non-SUSY attractor flow equations (\ref{firstU}) and
(\ref{firstz}). To show this, we first realize that the exact solutions for the moduli are obtained by replacing
the electric and magnetic charges by appropriate harmonic functions \cite{Kallosh:2006ib}
\begin{eqnarray}\label{attsolhar26}
z^{a}(\tau)=\pm i\frac{1}{H^{0}(\tau)H_{a}(\tau)}\sqrt{H^{0}(\tau)H_{1}(\tau)H_{2}(\tau)H_{3}(\tau)}\ .
\end{eqnarray}
Moreover, the solution for the metric field in the non-SUSY branch is given by \cite{Kallosh:2006ib}
\begin{eqnarray}\label{usol26}
e^{-2U(\tau)}=\sqrt{4\,H^{0}(\tau)H_{1}(\tau)H_{2}(\tau)H_{3}(\tau)}\ .
\end{eqnarray}
Having the exact solutions for moduli and for the metric field, we can construct the $R$-function (\ref{R26}) and
explicitly verify the first order equations (\ref{firstU}) and (\ref{firstz}). Therefore, we find
\begin{eqnarray}
&&\dot{U}(\tau)=-e^{U(\tau)}|R|=\frac{e^{-U(\tau)}}{4}\left[\frac{p^{0}}{H^{0}(\tau)}
+\frac{q_{a}}{H_{a}(\tau)}\right]\ .\label{1susy26}\\
&& \dot{z}^{I}(\tau)=-2e^{U(\tau)}g^{I\bar{J}}\partial_{\bar{J}}|R|=\frac{i}{4}
\frac{e^{-2U(\tau)}}{H^{0}(\tau)H_{I}(\tau)}\left[2\frac{q_{I}}{H_{I}(\tau)}-
\frac{q_{a}}{H_{a}(\tau)}+\frac{p^{0}}{H^{0}(\tau)}\right]\ ,\label{2susy26}
\end{eqnarray}
which shows perfect agreement between the two sides of the equalities in the first order differential equations of
motion.

\subsubsection{D0-D4 system}\label{D04sec}
The second example we consider is D0-D4 system. From (\ref{STUsup}), the superpotential for this system reads
\begin{eqnarray}\label{W04}
W(z^{1},z^{2},z^{3})=q_{0}+p^{1}z^{2}z^{3}+p^{2}z^{1}z^{3}+p^{3}z^{1}z^{2}\ .
\end{eqnarray}
The attractor values of the moduli in the non-SUSY branch have been previously found in
\cite{Tripathy:2005qp,Kallosh:2006bt} by minimizing the black hole potential $V_{BH}$ and they are given by
\begin{eqnarray}\label{z04}
z^{1}=\pm i\sqrt{\frac{q_{0}p^{1}}{p^{2}p^{3}}}\quad,\quad z^{2}=\pm
i\sqrt{\frac{q_{0}p^{2}}{p^{1}p^{3}}}\quad,\quad z^{3}=\pm i\sqrt{\frac{q_{0}p^{3}}{p^{1}p^{2}}}\ .
\end{eqnarray}
As in the previous example, the signs of the solutions are chosen such that the K\"{a}hler volume is positive.
\par
To construct the $R$-function for this example, we flip the signs of all D4-brane charges:
$p^{a}\rightarrow-p^{a}$ in the central charge. Under these sign flips, the duality invariant ${\cal
I}=\sqrt{4\,q_{0}p^{1}p^{2}p^{3}}$ is transformed to ${\cal I}=\sqrt{-4\,q_{0}p^{1}p^{2}p^{3}}$. The $R$-function
is then explicitly given by
\begin{eqnarray}\label{R04}
|R(z,\bar{z})|=\big|e^{K/2}(q_{0}-p^{1}z^{2}z^{3}-p^{2}z^{1}z^{3}-p^{3}z^{1}z^{2})\big|\ .
\end{eqnarray}
One can check that the critical points of the above $R$-function coincide with the non-SUSY attractor values for
moduli (\ref{z04}). As in the previous example, the exact non-SUSY solutions are obtained by replacing the charges
by harmonic functions in the attractor values of the moduli
\begin{eqnarray}\label{zsol04}
z^{a}(\tau)=\pm i\frac{H_{0}(\tau)H^{a}(\tau)}{\sqrt{H_{0}(\tau)H^{1}(\tau)H^{2}(\tau)H^{3}(\tau)}}\ .
\end{eqnarray}
Similarly, the exact non-SUSY solution for the metric field is given by
\begin{eqnarray}\label{U04}
e^{-2U(\tau)}=\sqrt{4\,H_{0}(\tau)H^{1}(\tau)H^{2}(\tau)H^{3}(\tau)}\ .
\end{eqnarray}
Now, we have the exact non-SUSY solution for all moduli and the metric field. Therefore, we can explicitly examine
the first order equations of motion (\ref{firstU}) and (\ref{firstz}) for this example. After some algebra, we
find
\begin{eqnarray}
&&\dot{U}(\tau)=-e^{U(\tau)}|R|=-\frac{1}{4}\left[\frac{q_{0}}{H_{0}(\tau)}
+\frac{p^{a}}{H^{a}(\tau)}\right]\ ,\label{udot04}\\
&& \dot{z}^{I}(\tau)=-2e^{U(\tau)}g^{I\bar{J}}\partial_{\bar{J}}|R|=i\,e^{2U(\tau)}H_{0}(\tau)
H^{I}(\tau)\left[\frac{q_{0}}{H_{0}(\tau)}+2\frac{p^{I}}{H^{I}(\tau)}- \frac{p^{a}}{H^{a}(\tau)}\right]\
,\label{zdot04}
\end{eqnarray}
which shows that the first order differential equations of motion are fulfilled. One can generalize the solutions
for the general case with arbitrary number of vector multiplets. For details see appendix \ref{app}.

\subsubsection{D0-D2-D4 system}

For this system, the superpotential can be written as
\begin{equation}
W=q_0 + q_a z^a + p^1(z^2 z^3) + p^2(z^1 z^3) + p^3(z^1 z^2)\ .
\end{equation}
The U-duality invariant is ${\cal I}= 4 q_0 D-(-q_1 p^1 + q_2 p^2 + q_3 p^3)$. Assume that the charge
configuration given is SUSY, so that ${\cal I}$ is positive. Then one can flip the sign of the invariant and go
towards the non-SUSY case by the following operations
\begin{eqnarray}
p^a&\rightarrow&-p^a\ ,\\
q_0&\rightarrow&q_0+\frac{1}{6}D^{ab}q_a q_b\ .
\end{eqnarray}
Here $D^{ab}=-\frac{3}{D}p^a p^b=(D_{ab})^{-1}$. These transformations are more complicated than those of the
axion-free systems described above. But we observe that under these transformations, the quantity
$\hat{q_0}=q_0+\frac{1}{12}D^{ab}q_a q_b$ is invariant. Also, from \cite{Tripathy:2005qp,Nampuri:2007gv}, we
observe that one can rewrite the superpotential as that of a D0-D4 system with a D0-brane charge $\hat{q_0}$ and
the same values of the D4-brane changes under the identification
\begin{eqnarray}
\hat{q_0}&=&q_0+\frac{1}{12}D^{ab}q_a q_b\ ,\\
\hat{z}^a&=&z^a-\frac{1}{6}D^{ab}q_b\ .
\end{eqnarray}
Here the hatted charges and moduli correspond to those of the equivalent D0-D4 system.
 The U-duality invariant when written in terms of $\hat{q_0}$ is $4 \hat{q_0} D$. This  is precisely the invariant
 of the D0-D4 system. Hence we can work in the D0-D4 'frame' and construct the $R$-function as described in
 section \ref{D04sec}. As shown there, the extremization of the $R$-function give the attractor values of the
 moduli- as
 \begin{equation}
 z^a= \pm i p^a\sqrt{-\frac{\hat{q_0}}{D}}\ .
 \end{equation}
Reverse-transforming the moduli to the D0-D2-D4 configuration, we obtain the attractor values in the D0-D2-D4
system as
 \begin{equation}
 z^a= \pm i p^a\sqrt{-\frac{\hat{q_0}}{D}}+\frac{1}{6}D^{ab}q_a q_b\ .
 \end{equation}
 The $R$-function constructed in the D0-D4 case can be explicitly written down in the D0-D2-D4 coordinates to give
 the corresponding $R$-function.

\subsubsection{D0-D6 system}
The superpotential in this case is
\begin{equation}
W= q_0-p^0 z^1 z^2 z^3\ .
\end{equation}
The U-duality invariant is ${\cal I}= 4 (q_0 p^0)^2$. Hence the invariant is positive definite and no real-valued
charge transformations which preserve the absolute value of the invariant can flip its sign. This system unlike
the previous ones has no SUSY equivalent. So the simplest way to construct $|R|$ is to dualize this system to the non-SUSY D0-D4 system and then construct $|R|$ for this system . The attractor values for the moduli so found can then be rewritten in D0-D6 language, as can the entropy.

\subsection{General rules for attractors}
For the SUSY case, the general rule for finding the attractor values is very clear. One simply minimizes the
absolute value of the central charge w.r.t. to the moduli and this gives the attractor values. For the non-SUSY
case, we observe that for every non-SUSY black hole, there is a SUSY black hole with the same mass
 and entropy saturating the BPS bound. Calling $|Z|_s$ the central charge of the SUSY black hole, one sees that
 the mass $M$ of the non-SUSY black hole must satisfy $M=|Z|_s$ and so $|Z|_s$ is potentially a candidate for the
 $R$-function. We can check this claim by solving $\partial_a|Z|_s=0$ to obtain the non-SUSY attractor values
 which must be exactly the same as the attractor values of the corresponding SUSY black hole. Explicit
 calculations demonstrating this for the STU model have been done in the preceding subsection while calculations
 for a general prepotential are done for the D0-D4 case in the appendix.
One reasonable worry is the invariance of the potential under $|Z|\rightarrow |Z|_s$ via charge-flips. One can
explicitly show for various charge configurations that the potential remains invariant under this transformation.
Hence the $R$-function for the non-SUSY black hole of the first class is $M=|Z|_s$.
Hence the general rule for extremal blackholes which display the attractor mechanism is that the mass function $M=|R|$ gets extremized along the attractor flow from asymptotic infinity to the horizon. Extremizing the mass function is equivalent to solving for $\partial |Z_s|=0$ and this gives the attractor value of the moduli. For the purpose of the present discussion, we hereafter refer to the charge lattice of the given black hole by $\Gamma$ and the transformed charge lattice  in $R$ as $\Gamma_s$. For SUSY configurations, $\Gamma=\Gamma_s$.
\par
The previous sections show that the attractor mechanism for SUSY and non-SUSY extremal black holes are on the same footing, thus rendering all the technology available for study of SUGRA SUSY configurations amenable to being applied to the non-SUSY configurations. One of the most exciting developments in the SUSY attractor story is the development of split attractors in \cite{Bates:2003vx} and \cite{Denef:2000nb}. This technology uses the first order attractor formalism and can now be extended to non-SUSY configurations too as we discuss in the subsequent sections.
\section{Split Attractor Flows}
The attractor equations in SUGRA for extremal black holes fix the scalar moduli at the horizon and make the mass of the black hole flow from its ADM mass at asymptotic infinity (determined completely by the values of the scalar fields at infinity and the charges of the black hole) to its value at the horizon (determined completely in terms of the charges of the black hole). As one movies in asymptotic moduli space, along certain loci referred to as lines of marginal stability,  the ADM mass becomes equal to the sum of the ADM masses of two separate charge configurations, and then as each of these loci are crossed the ADM mass becomes smaller than the sum of the ADM masses of the two configurations. In these regions of moduli space, its possible for the single center black hole to exist as a non-spherically symmetric two-centered configuration\footnote{Such configurations were first analysed in \cite{Galli:2009bj}.}. The distance between the two centers diverges at the line of marginal stability (LMS) and on one side of the LMS, it is positive and fixed in terms of the Dirac angular momentum between the two centers. Since, for an extremal black hole, $M=|R|$, the two charge configurations by indices '1' and '2' , the LMS is defined by the condition,
\begin{equation}\label{splitcond1}M = M_1 + M_2\ .\end{equation}
Using $M=|R|$, this can be rewritten as
\begin{equation}
\label{splitcond11}
|R| = |R_1| + |R_2|\ .
\end{equation}
This implies, in turn, a condition on the charge splits
\begin{equation}\label{splitcond2}
\Gamma_s = \Gamma^1_s + \Gamma^2_s\ .\end{equation}Notice that (\ref{splitcond2}) must hold all over moduli space where the two-centered system exists, since charge is a constant.
%%%%%%%%%%%%%%%%%%%%%%%%%%%%%%%%%%%%%%%%%%%%%%%%%%%%%%%%%%%%%%%%%%%%%%%%%%%%%
\begin{figure}[ht]
\begin{center}
\SetScale{1}
    \begin{picture}(300,160)(0,0)
    \ArrowLine(140,115)(140,50)
    \DashLine(140,148)(140,115){4}
    \CCirc(140,150){2}{Black}{Gray}
    \CCirc(95,5){2}{Black}{Gray}
    \CCirc(185,5){2}{Black}{Gray}
    \ArrowArc(140,5)(45,90,180)
    \ArrowArcn(140,5)(45,90,0)
    \Text(102,150)[l]{$\tau=0$}
    \Text(149,85)[l]{$M(\Gamma)$}
    \Text(52,6)[l]{$\tau_{1}=\infty$}
    \Text(192,6)[l]{$\tau_{2}=\infty$}
    \Text(177,40)[l]{$M_{2}(\Gamma_{2})$}
    \Text(67,40)[l]{$M_{1}(\Gamma_{1})$}
    \DashCArc(140,50)(6,0,360){2}
    \LongArrowArcn(185,5)(64,130,90)
    \Text(189,68)[l]{$M=M_{1}+M_{2}$}
    \end{picture}
\end{center}
\caption{The fork flow of the double-center solution. Mass conservation applies necessarily at the fork.}
\end{figure}
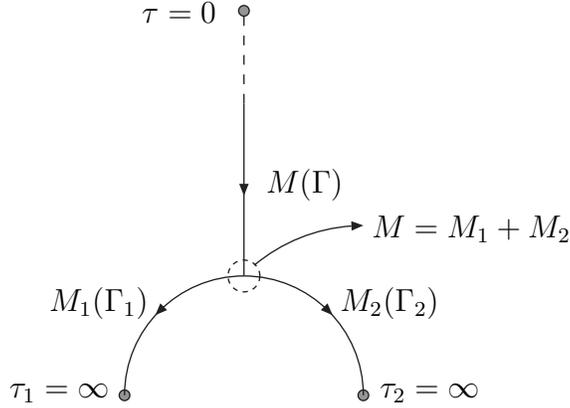
%%%%%%%%%%%%%%%%%%%%%%%%%%%%%%%%%%%%%%%%%%%%%%%%%%%%%%%%%%%%%%%%%%%%%%%%%%%%%

We now give a short review of the split attractor flow for two-centered SUSY configurations ala \cite{Denef:2000nb} and \cite{Bates:2003vx}.
\subsection{SUSY split attractor flows}
The linear algebraic equations that solve for the first order attractor equations for a SUSY black hole in an ${\cal N}=2$ theory are
\begin{eqnarray}
2 e^{-U}\, \Im(e^{-i \alpha} X^I)&=& H^I\ ,\\
2 e^{-U}\, \Im(e^{-i \alpha} F_I)&=& H_I\ .
\end{eqnarray}
All notation above and hereafter follows \cite{Bates:2003vx}.
In the above equations, $\alpha$ denotes the argument of the central charge. The $H$'s are harmonic functions of the electric and magnetic charges of the form
\begin{eqnarray}
H^I&=&h^I+\Gamma^I \tau\ ,\\
H_I&=&h_I+\Gamma_I \tau\ .
\end{eqnarray}
Here, as before, $\tau$ is the inverse of the radial coordinate.
One can derive these equations by considering the 1D effective Lagrangian density and writing it as a perfect square in special geometry notation as $|(\partial_\tau + \dot{Q}+ \dot{\alpha})(e^{-U}e^{-i \alpha}\Omega)-\Gamma |^2$. Here $Q=\Im(\partial_a K dz^a)$. Here $\alpha$ is the argument of the function $R$ that appears in the 1D effective potential as $V_{BH}=|R|^2+4 g^{a\bar{b}}\partial_a|R| \partial_{\bar{b}}|R|$. One can then show $ \dot{Q}+ \dot{\alpha}=0$ and hence arrive at the linear algebraic equations which solve the resulting first order PDE. For the SUSY case, $R=Z$ and hence $\alpha =\mbox{arg}(Z)$. Denoting the coordinate distance between the two centers by $a$, we can write the algebraic equations for the two centered splits in the same form as the single center flow except that the harmonics become
\begin{eqnarray}
H^I&=&h^I+\frac{{\Gamma^1}^I}{r} + \frac{{\Gamma^2}^I}{|r-a|}\ ,\\
H_I&=&h_I+\frac{\Gamma^1_I}{r}  + \frac{\Gamma^2_I}{|r-a|}\ .
\end{eqnarray}

At all points in moduli space, $\Gamma = \Gamma^1 + \Gamma^2$ as seen from (\ref{splitcond2}) and $Z=Z_1 + Z_2$.
The LMS is defined by (\ref{splitcond1})  which in this case turns out to be $|Z|=|Z_1|+|Z_2|$. In other words, the LMS is defined by the locus of phase alignment of the two constituents with the phase of the net central charge.
An integrability criteria involving the Harmonics - $H^I{\nabla}^2H_I-H_I{\nabla}^2H^I=0$ gives the expression for the distance at any point in asymptotic moduli space
\begin{equation}
a=-\frac{\langle\Gamma^1,\Gamma^2\rangle}{2\, \Im (e^{-i \alpha} Z_1)}=-\frac{q^1_I p_2^I-q^2_I p_1^I}{q^1_I h^I- p_1^I h_I}\ .
\end{equation}
The region of existence of the two-centers begins at the LMS where the denominator vanishes and the distance between the two centers diverges. To one side of the LMS, the phase of $Z_1$ exceeds that of $Z$ and the distance becomes negative indicating non-existence of two-centered solutions while to the other side, it becomes less than that of the $Z$ giving a positive coordinate distance and hence a physically sensible solution. In this region, $|Z|<|Z_1|+|Z_2|$ implying $M < M_1 + M_2$ which is the usual necessary condition for bound state formation of two particles.
\subsection{Non-SUSY split attractor flows}
In the non-SUSY case, we can draw precise parallels with the SUSY formalism. The $R$ function appearing in the Lagrangian is no longer the central charge but $M=|R|$ is still satisfied as in the SUSY case. The 1D Lagrangian density can be written as a perfect square just as in the SUSY case whilst replacing $Z$ by $R$ and $\Gamma$ by $\Gamma_s$-
\begin{equation}
{\cal L}=|(\partial_\tau + \dot{Q}+ \dot{\alpha})(e^{-U}e^{-i \alpha}\Omega)-\Gamma_s |^2\ .
\end{equation}
Here, $\alpha$ is the argument of $R$.
The same reasoning as in the SUSY case leads to an identical algebraic equation series.
Applying the integrability criteria gives the distance between the two centers to be
\begin{equation}
a=-\frac{\langle\Gamma_s^1,\Gamma_s^2\rangle}{2\,\Im (e^{-i \alpha} R_1)}=-\frac{{q_s^1}_I {{p_s}_2}^I-{q_s^2}_I {{p_s}_1}^I}{{q_s^1}_I h^I- {{p_s}_1}^I h_I}\ .
\end{equation}
The LMS criterion is defined by $|R|=|R_1|+|R_2|$ and the region of existence of the two centers is defined by $|R|<|R_1|+|R_2|$ with the constraints $\Gamma_s = \Gamma^1_s + \Gamma^2_s$ and $R=R_1 + R_2$.
However, as we shall see below, the non-SUSY two-centered case can exist only when the distance between the centers is not fixed.

\section{Interpretation and Discussion}
We now see that using the first order formalism essentially makes the non-SUSY split flow problem as tractable as the SUSY one. In both cases, this formalism gives the answer to the question of whether a given parent charge configuration can split into two constituents with precisely defined charges at a certain point in asymptotic moduli space. In the SUSY case, by construction of the split problem the constraints $\Gamma = \Gamma_1 + \Gamma_2$ and $Z = Z_1 + Z_2$ is directly imposed. One can use the integrability criterion and the linear addition of the harmonic functions of the two constituents to find the distance between the two centers\footnote{For recent work on split configurations and wall-crossing, see \cite{Andriyash:2010qv,Andriyash:2010yf}.}. In the non-SUSY case however, the split constraints $\Gamma_s = \Gamma^1_s + \Gamma^2_s$ and $R=R_1 + R_2$ are obtained via the construction of the $R$ function from the $Z$ function by appropriate transformation on charges taking $\Gamma$ to $\Gamma_s$ and they are {\it not} concordant with the SUSY constraints, $\Gamma = \Gamma_1 + \Gamma_2$ and $Z = Z_1 + Z_2$, simply indicating that two SUSY constituents cannot form a SUSY as well as a non-SUSY bound state in the same region of moduli space. Hence, in the non-SUSY case, given a parent charge configuration and a potential split, one converts the charges to the $\Gamma_s$ form for both parent and daughter configurations and then checks if $\Gamma_s = \Gamma^1_s + \Gamma^2_s$ is satisfied. Only if this is so can the split occur. Now, generically, one goes from the $\Gamma$ to the $\Gamma_s$ frame by a select transformation of the charges. If the transformation matrix is diagonal then and only then can the non-SUSY constraints and the charge conservation constraint be met. However in this case, the parent and the daughter must all have the same charges and the Dirac angular momentum between the daughter elements, which fixes the distance between the centers, vanishes. This effectively leaves the distance between the two centers unfixed and unstable to perturbation. Hence they exist only at the LMS.\footnote{This was first pointed out in \cite{Galli:2009bj} and mentioned recently in \cite{Ferrara:2010ru}.}
\par
Physically, one can see that this feature of non-SUSY split centers arises because of the fact that the central charges of the daughter constituents always add up as complex numbers to the parent central charge. The central charge is linear in charges and moduli while for any daughter elements with a Dirac angular momentum-fixed distance, the mass conservation condition will never hold if the central charge conservation condition does since the mass given by $|R|$ is not linear in charges. This is simply an indication of the fact that non-BPS objects which are bound have a binding energy that is always non-zero and this absence of a threshold is what renders stable non-SUSY split configurations untenable. For a detailed explication of the split flow formalisms in the 5D uplifts of the 4D non-supersymmetric charge configurations, treated using the second-order equations, refer to \cite{Bena:2009qv,Bena:2009fi,Bena:2009en,Bobev:2009kn}.

Now the prescription for going to the $\Gamma_s$ frame or alternatively, from the $Z$ to the $R$ frame is not unique. For example in the D0-D4 case, one can either flip the signs of the three D4 branes or the D0 brane or one of the D4 branes. All these give different $R$ functions with the same extremum value of $|R|$. Hence, at the extremum, or the attractor point, $R$ is undefined up to a phase. Remembering that $R$ is the central charge of the equivalent SUSY system, we see that this ambiguity at the attractor point is essentially the $U(1)_R$ ambiguity of choosing a basic in supercharge space in ${\cal N}=2$ theories.
Hence fixing a point in this automorphism group fixes the $R$-function precisely.\footnote{This in fact is the reason behind labeling it the $R$ function.}

\section{Summary of Results in Attractors}
The 1D effective Lagrangian can be thought of as the motion of a non-relativistic particle in the $(2N_v + 1)
(U,z^a,\bar{z^a})$ space. The geodesics of the particle are attractor flows under the extremality constraints on
the Hamiltonian, and the potential can be written in terms of the mass of the black hole as
$V_{BH}=4g^{a\bar{b}}\partial_{a}M\partial_{\bar{b}}M+M^{2}$. One can then use the Bogomol'yni squaring technique to
write down the first order equations for the moduli and the metric field. The mass is equal to the central charge
for SUSY, and is equal to the central charge
of the corresponding SUSY equivalent if it exists. Note that for non-SUSY systems, the Bogomol'yni trick gives the
extremal but non-BPS bound. This successfully sets up the first order framework of the attractor mechanism on the
same footing for SUSY and non-SUSY black holes and essentially, opens up the way to parallel developments in SUGRA
BPS black holes for SUGRA non-SUSY black holes. However the presence of the attractor mechanism throws up the
conundrum of the independence of the entropy from the asymptotic moduli especially the dilaton. In certain special
cases, this lends itself to an explanation. Firstly note that, as shown in \cite{Ceresole:2007rq}, 5D
supersymmetric black holes give rise to a supersymmetric and a non-supersymmetric branch in 4D, under the 4D-5D
lift. This implies that the $\partial{|Z|}=0$ condition in 5D leads to both a $\partial{|Z|}=0$ and a
$\partial{R}=0$ condition in 4D. We use this fact in the subsequent discussion.
\par
Consider, a 4d black hole in a N=4 heterotic compactification on $T^4\times S^1 \times \tilde{S^1}$ or its dual
IIA compactification on $K3\times S^1 \times \tilde{S^1}$. We write the heterotic charges in the IIA language here
and designate them as electric or magnetic in the heterotic frame. We restrict ourself to a STU duality subgroup
of the N=4 theory with charges occupying a SO(2,2) Narain lattice. The generic electric and magnetic charge vector
in this case is written in standard notation as $Q=(q_0, -p^1, q_2,q_3)$ and $P=(q_1, p^0,p^3,p^2)$. We restrict
ourself for ease of conceptual understanding to 4-charge systems with non-zero $q_1, p^0, q_2$ and $q_3$. Then the
4d duality invariant is given by $I_4= 4 p^0 q_1 q_2 q_3$. The IIA charges can be lifted up to 5d with the 5d
duality invariant being $I_4= q_1 q_2 q_3$. The (D1-D5-P-KK) system which is the canonical system considered for
counting SUSY dyons in 4d N=4 string theories can be a $(q_2-q_3-(-q_1)-(-p^0))$ in the IIA frame and this system
can be thought of (at an appropriate point) as a D1-D5-P system at the center of Taub-Nut where its basically a 5d
three-charge black hole with an duality invariant that is the product of the three charge-a fact that is fully
consistent with the identification of three charges with the three charges of the IIA and the 5d duality invariant
defined above. Now while counting the D1-D5-P-KK system, we consider the Hilbert space as the direct product of
three different parts- one coming from the internal SCFT of the D1-D5-P, a second arising from the bound states of
the KK-P and a third arising from the motion of the D1-D5-P system in the Taub-Nut space. Notice that, from our
identification of 4d and 5d charges, the sign of the 5d duality invariant and its BPS condition is independent of
the sign of the KK monopole. So we choose a sign of the KK monopole which is 4d non-BPS and then count the entropy
of the D1-D5-P system, with Q1, Q5 and P of the same sign, and given by an elliptic genus which asymptotically
gives rise to the Bekenstein Hawking entropy. Now assuming the KK monopole to have unit magnitude, the
corresponding 4d black hole has the same large charge entropy. A 5d non-BPS black hole configuration with unit
magnitude of the KK monopole would have $q_1, q_2 < 0$ and $q_3 > 0$. The charge vectors would be in this case,
$(Q|P)=(0,0,q_2,q_3|q_1,-1,0,0)$. Here, each hyperbolic lattice corresponds to one of the two circles of the
compact manifold modulo the $T^2$ in the heterotic frame. Flipping the two circles and then performing an
electric-magnetic $Q\rightarrow P, P\rightarrow-Q$ transformation, we get $(Q|P)=(-q_2,-q_3,0,0|0,0,-q_1,1)$. Thus
we end up with a configuration that corresponds to a negative value of the KK monopole, and positive values of the
other three charges, which ends being a 4d non-BPS but a 5d-BPS. The large charge entropy remains unchanged here.
Thus the asymptotic large charge entropy of a 4d non-BPS black hole which derives from a 5d BPS black hole and
hence from the related elliptic genus can be related to a 4d non-BPS black hole deriving from a 5d non-BPS black
hole. Both classes of black holes can be related by U-duality to any other 4d non-BPS black hole and hence one has
finally a microscopic understanding of why the Bekenstein Hawking entropy computation is effectively renormalized
as one goes from weak to strong 4d string coupling
\par
The non-renormalizability of the entropy with coupling strength can be rigorously explained for certain classes of
supersymmetric black holes where the entropy is related to an index computation. But for non-supersymmetric black
holes, one may  try to explain it by looking at the strong coupling side- namely the underlying CFT. This
motivates the ensuing discussion on attractor flows in CFT.

\section{Attractor Moduli Flow in CFT}
In this section, we first provide the brief background in CFT necessary to introduce moduli flow equations in CFT. This closely follows the chain of logic and notation of \cite{Brunner:2010xm}. In the next subsection, we proceed to make a formal correspondence between the moduli flows in CFT and the attractor flows in SUGRA.  In the subsequent subsection, we discuss the implications of this correspondence to attractors flows and forks and to the non-renormalization of the Bekenstein-Hawking entropy, in line with all evidence for its universality in the large charge limit of solitonic states.
\subsection{Background on CFT moduli flow}
As has been pointed out in \cite{Bachas:2007td} and explored in \cite{Brunner:2008fa} and \cite{Brunner:2007ur}, to each bulk perturbation of a CFT, one can associate a unique defect line
between the IR and UV theories of the corresponding RG flow. There exists a Casimir energy between defect lines
and boundary conditions in CFT (See \cite{Bachas:2001vj}.) and a consequent attractive or repulsive force between them which seeks to
extremize this energy. Hence a defect line or boundary condition in CFT defines a tangent vector to the
deformation space of this CFT, by requiring that the energy gain is maximal in this direction. If the defect or
boundary condition can be deformed smoothly under bulk deformations to any point in moduli space, then the tangent
vectors form a vector field and give rise to a flow in moduli space given by
\begin{equation}
\frac{d\lambda^{i}}{dt}=-g^{ij}\partial_{j}E_{BD}\ .
\end{equation}
Here, $E_{BD}$ is the Casimir energy between the boundary condition $B$ and the deformation defect $D$ which
produces a smooth bulk deformation of B in moduli space,while $g_{ij}$ is the Zamolodchikov metric, and
$\lambda^{i}$ are the moduli fields. As shown in \cite{Brunner:2010xm}, the flow of the gradient of the Casimir
energy between a boundary condition and a defect line is a constant re-parametrization of the gradient flow of the
logarithm of the $g$-function associated with the defect line. This allows a rewriting of the moduli flow
equations as
\begin{eqnarray}\label{CFTeq}
\frac{d\lambda^{i}}{dt}&=&-g^{ij}\partial_{j}\log(g)\ .
\end{eqnarray}
 Now, the mass $M$ of a particle in CFT is the $g$-function of its corresponding boundary condition. So, one can
 as well write the moduli flow equations as
 \begin{eqnarray}\label{CFTeqM}
\frac{d\lambda^{i}}{dt}&=&-g^{ij}\partial_{j}\log(M)\ .
\end{eqnarray}
For a BPS particle, $M=|Z|$ and the flow equations becomes
\begin{eqnarray}\label{CFTeqZ}
\frac{d\lambda^{i}}{dt}&=&-g^{ij}\partial_{j}\log(|Z|)\ .
\end{eqnarray}
\subsection{CFT/SUGRA correspondence}
Now, to establish a correspondence between the CFT flow equations and the first order attractor equations in
SUGRA
\begin{eqnarray}
\dot{U}&=&-e^{U} |R|\ ,\label{attr1}\\
\dot{z^a}&=&-2 e^{U} g^{a\bar{b}}\partial_{\bar{b}}|R|\ .\label{attr2}
\end{eqnarray}
We substitute for ${e^{U}}$ from (\ref{attr1}) in (\ref{attr2}) to obtain
\begin{equation}
\dot{z^a}= 2 \dot{U} g^{a\bar{b}}\frac{\partial_{\bar{b}}|R|}{|R|}\ . \label{attr3}
\end{equation}
The derivative in (\ref{attr3}) is w.r.t. $\tau$, the inverse of the radial coordinate in SUGRA. Denoting $t=-2 U(\tau)$,
we obtain the CFT moduli flow equations-
\begin{eqnarray}
\frac{d\lambda^{i}}{dt}&=&-g^{ij}\partial_{j}\log(M)\ ,
\end{eqnarray}
where the $\lambda$ stands for the moduli represented by $z$ in SUGRA, the Zamolodchikov metric is identified with
the K\"{a}hler metric in SUGRA, and the $R$-function is identified with the mass function- this is consistent
with identifying the $R$-function with the mass of the black hole in SUGRA and thinking of its flow from asymptotic
infinity to the horizon as a flow from the ADM mass to the Bertotti-Robinson mass in the near horizon geometry.
Notice that the of the two attractor equations in the bulk- only the one corresponding to moduli flow has a CFT
correspondence. The metric field is simply a re-parametrization of the radial direction which acts as an energy
scale for the RG flow in the CFT. Hence the CFT flow does not include information about the dynamics of the metric
field. Finally, for the supersymmetric state, $M=|Z|$, and we obtain the moduli flow associated with a
supersymmetric boundary condition.
\subsection{Discussion and implications}
The full import of the correspondence made in the previous section lies in its application to the phenomenon of attractors flows. In SUGRA, the attractor flow is a directed trajectory from asymptotic moduli space $(h_I, h^I)$ to the near-horizon geometry. In ${\cal N}=2$ SUGRA, non-trivial SUSY solutions in a bosonic background must satisfy the following integrability criterion-$H_I \nabla^2 H^I- H^I\nabla^2 H_I=0.$ This reduces to $\langle \mathbf{\Gamma},\mathbf{h}\rangle=0$. The inner product is defined by the symplectic structure on the cohomology of the Calabi-Yau.
At points in asymptotic moduli space where the absolute value of the central charge is the sum of the absolute values of central charges of two charge configurations, the attractor flow forks into two flows corresponding to the two centers. This situation occurs if the phases of the two constituents are aligned, and the loci at which this happens are the LMS.\footnote{We consider only single centers with $|Z|\neq 0$ everywhere.}The regions of existence of the single center and multi-centers in the space defined by the phase difference $\delta\alpha$ between the central charges is shown in Fig.2. The Euclidean coordinate distance between the two centers ranges from infinity at the LMS to a finite distance $a$ between the centers at points in moduli space away from the LMS given by
\begin{equation}
\frac{a}{\langle \Gamma_1,\Gamma_2\rangle}=\frac{|Z_1 + Z_2|}{2\,\Im(Z_1\bar{Z_2})}
\end{equation}
The distance is positive definite and finite on one side of the LMS and negative on the other side indicating non-existence of the two centered system.
%%%%%%%%%%%%%%%%%%%%%%%%%%%%%%%%%%%%%%%%%%%%%%%%%%%%%%%%%%%%%%%%%%%%%%%%
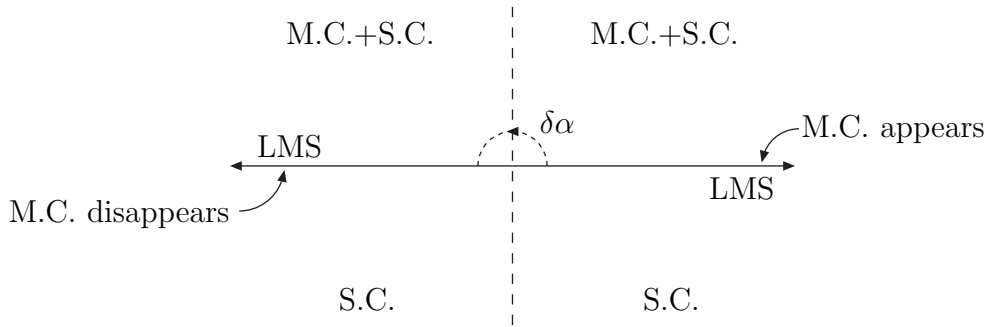
\begin{figure}[ht]
\begin{center}
\SetScale{1}
    \begin{picture}(300,140)(0,0)
    \LongArrow(150,70)(255,70)
    \LongArrow(150,70)(45,70)
    \DashLine(150,10)(150,130){4}
    \DashArrowArc(150,70)(13,0,180){2}
    \Text(225,62)[l]{LMS}
    \Text(54,78)[l]{LMS}
    \Text(161,87)[l]{$\delta\alpha$}
    \Text(200,20)[l]{S.C.}
    \Text(85,20)[l]{S.C.}
    \Text(65,120)[l]{M.C.+S.C.}
    \Text(180,120)[l]{M.C.+S.C.}
    \Text(260,84)[l]{M.C. appears}
    \Text(-40,52)[l]{M.C. disappears}
    \LongArrowArc(258,70)(14,90,175)
    \LongArrowArc(47,70)(17,270,355)
    \end{picture}
\end{center}
\caption{Regions of existence of the single-centered (S.C.) and multi-centered (M.C.) solutions in an abstract 2D space with angular coordinate $\delta\alpha$, which is the phase difference between the central charges of the two centers.}
\end{figure}
%%%%%%%%%%%%%%%%%%%%%%%%%%%%%%%%%%%%%%%%%%%%%%%%%%%%%%%%%%%%%%%%%%%%%%%%

On the gravity side of the story, the natural dynamical formalism to study attractor flows is that of the black hole potential $V_{BH}$ generating a second order equation for moduli flow. The attractor mechanism allows us to write these equations as first order ones.
Now in the CFT, the moduli flow equations were derived purely from a RG perspective by studying the beta function of the moduli as the CFT flowed from UV to IR under a perturbation generated by a boundary condition like a D-brane. The presence of the IR fixed point renders this flow to be a first order equation naturally. But since this system is the underlying description of the attractor flow induced by a black hole in moduli space in the strongly coupled CFT regime, we should naturally be able to physically argue for and derive the existence of the effective potential governing the dynamics of the attractor flow.\footnote{Note that the potential so derived will be generically different from the one obtained in SUGRA and will be the full quantum corrected version of the SUGRA $V_{BH}$.}
\par
We first think of the moduli flow as generating trajectories in moduli space characterized by an affine parameter $t$. The flow terminates on the boundary condition represented by the attractor point at $t=0$. We now (with prescient foresight) define a 'double extremal' trajectory as one that takes a constant value all the way from $t=\infty$ to $t=0$ and take this as the 'canonical geodesic'.
Any other moduli flow in this space terminating at the same attractor point suffers a geodesic deviation w.r.t. the canonical trajectory. This geodesic deviation is experienced as an acceleration generated by a potential for a particle moving along this trajectory.
 %%%%%%%%%%%%%%%%%%%%%%%%%%%%%%%%%%%%%%%%%%%%%%%%%%%%%%%%%%%%%%%%%%%%%%%%
\begin{figure}[ht]
\begin{center}
\SetScale{1}
    \begin{picture}(300,140)(0,0)
    \ArrowLine(110,70)(250,70)
    \DashLine(30,70)(110,70){4}
    \CCirc(30,70){2}{Black}{Gray}
    \CCirc(250,70){2}{Black}{Gray}
    \ArrowArcn(171,-20)(120,85,50)
    \Line(110,104)(182,100)
    \DashLine(30,108)(110,104){4}
    \CCirc(30,108){2}{Black}{Gray}
    \ArrowArc(171,160)(120,275,310)
    \Line(110,36)(182,40)
    \DashLine(30,32)(110,36){4}
    \CCirc(30,32){2}{Black}{Gray}
    \DashLine(30,140)(30,0){4}
    \Text(33,130)[l]{$t=\infty$}
    \Text(258,70)[l]{$t=0$}
    \Text(113,79)[l]{$\lambda^{i}_{c}(t)$}
    \Text(113,113)[l]{$\lambda^{i}_{1}(t)$}
    \Text(113,47)[l]{$\lambda^{i}_{2}(t)$}
    \end{picture}
\end{center}
\caption{Attractor flows from different asymptotic moduli or starting points. The middle trajectory is the double extremal trajectory which is the canonical geodesic w.r.t. the other two trajectories suffer geodesic deviation.}
\end{figure}
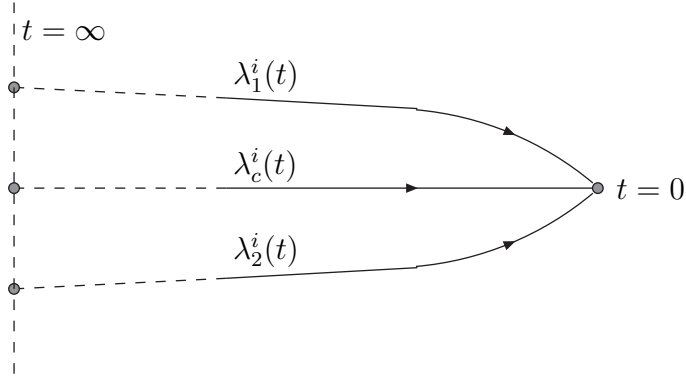
%%%%%%%%%%%%%%%%%%%%%%%%%%%%%%%%%%%%%%%%%%%%%%%%%%%%%%%%%%%%%%%%%%%%%%%%

 In order to derive the deviation from geodesic motion for the particle, we first choose a re-parametrization of the trajectory in terms of a positive definite variable $\tau$ and make the identifications-$t=-2 U(\tau)$ to obtain the two first order equations-
\begin{equation}
\dot{U}=-e^{U}|Z|\qquad,\qquad
\dot{\lambda}^i=-2e^{U}g^{rs}\partial_s|Z|.
\end{equation}
The independent variable in the above equations is the new affine parameter $\tau$.
These are formally the same as the attractor flow equations in SUGRA and as has been shown in section (3) satisfy the second order equations (\ref{attract1}) and (\ref{attract2}). The latter equation-
\begin{equation}
\ddot{\lambda}^i+\Gamma^i_{jk}\dot{\lambda}^j\dot{\lambda}^k=
e^{2U}g^{ij}\partial_j(|Z|^2+2g^{ij}\partial_i|Z|\partial_j|Z|),
\end{equation}
is a geodesic deviation equation for an accelerated particle with the acceleration generated by a potential $V_{\lambda}=\partial_j(|Z|^2+2g^{ij}\partial_i|Z|\partial_j|Z|)$. This is seen to be the same as the black hole potential $V_{BH}$. A vanishing RHS of the equation gives the geodesic equation in moduli space corresponding to an unaccelerated particle. The particle thus moves in a potential field from $\tau=0$ towards its extremum at the attractor point ($\tau=\infty$) where $\partial_jV_{\lambda}=0$.
This potential is an invariant under transformations which preserve the symplectic structure of the phase space of the particle, namely the moduli space. It is generally a function of both the moduli and the charges, and only at its extremum, where all moduli get fixed in terms of the charges does it become an invariant constructed purely out of the charges of the theory. In the limit of large charges this invariant is unique and is equal to the Bekenstein-Hawking entropy of the system. For the SUSY case, the extremum is defined by $\partial_j|Z|=0$ implying that the extremum value of the potential is purely the square of the mass of the BPS object. For non-SUSY configurations, the whole argument as presented above goes through with $|Z|$ replaced by the mass function $|R|$. The potential in this case is the expected $V_{\lambda}=V_{BH}=\partial_j(|R|^2+2g^{ij}\partial_i|R|\partial_j|R|)$. Hence we find that at the attractor point the extremum of the mass function is an invariant fixed purely in terms of charges and in the large charge limit yields the Bekenstein-Hawking entropy, indicating a non-re-normalization of the entropy even in non-SUSY cases.
\par
For multi-centered configurations in CFT, a forked flow equation just as in gravity is highly indicated and since the concept of geometrical radial distance is absent in the CFT, one way of measuring the 'distance' between the two centers is to find the geodesic deviation between the double-extremal trajectory of one of the centers.
The deviation is zero at all points in moduli space where only the single center exists and at the LMS and non-zero in regions where the multi-centered system exists. The CFT equations for moduli flow can be used to show that at the LMS, $|Z|=|Z_1|+|Z_2|$ and $\dot{\lambda}^i|Z|=\dot{\lambda}_1^i|Z_1|+\dot{\lambda}_2^i|Z_2|$.

\section{Conclusions}
We begin with the fact that the formalisms for the first order moduli flow in supersymmetric and non-supersymmetric black
hole backgrounds can be viewed on the same footing. We then, arrive at a solution generating technique for moduli flow
in non-supersymmetric black holes from those in the supersymmetric case. We establish an exact correspondence
between the attractor flow in gravity and moduli flow equations in the underlying CFT. We use these results to
indicate non-renormalization of the leading order entropy. In sum, the attractor formalism and its link to moduli flow in CFT can, on further exploration, yield, insights into non-supersymmetric single and multi-center solutions from both a gravitational and a CFT viewpoint. 

\vspace{1.1cm} \noindent {\bf Acknowledgements}
\vspace{0.3cm}\\
We take great pleasure in expressing our gratitude to Marco Baumgartl, Ilka Brunner, Gabriel Cardoso, Michael Haack, Renata Kallosh and Sandip Trivedi for informative and illuminating discussions and comments.
M.S. would like to thank the hospitality of the 8-th Simons Workshop in Mathematics and Physics where part of this work was carried out. S.N. is supported in part by the European Community's human potential program under contract
MRTN-CT-2004-005104 ``Constituents, Fundamental Forces and Symmetries of the Universe'' the Excellence Cluster,
``The Origin and the Structure of the Universe'' in Munich, and the German Research Foundation (DFG) within the
Emmy-Noether-Program (grant number: HA 3448/3-1). The work of M.S. is supported by a EURYI award of the European
Science Foundation.

\appendix
\section{Non-SUSY Solutions for General D0-D4 System}\label{app}
As we saw in section \ref{D04sec}, the attractor values  for the moduli of the STU model were constructed from the
corresponding $R$-function, and the known flow solutions for the moduli derived in \cite{Kallosh:2006ib} were
shown to satisfy the first order non-SUSY attractor equations. In this section, we generalize this result for an
arbitrary number of vector multiplets arising from D0-D4  compactifications on general Calabi-Yau manifolds. The
superpotential of the D0-D4 system with $N$ complex scalar fields, $z^{a}\ ,\ a=1,2,\cdots, N$, is given by
\begin{eqnarray}\label{W04app}
W(z^{a})=q_{0}-3\,d_{abc}\,p^{a}z^{b}z^{c}\ .
\end{eqnarray}
To construct the $R$-function, we apply the same rule as the STU example, namely flipping the sign of D4-brane
charges $p^{a}\rightarrow-p^{a}$. The corresponding $R$-function is then given by
\begin{eqnarray}\label{R04app}
|R(z,\bar{z})|=\big|e^{K/2}(q_{0}+3\,d_{abc}\,p^{a}z^{b}z^{c})\big|\ .
\end{eqnarray}
To find the attractor values of the moduli, we minimize the above $R$-function. Noting that the solutions are
axion-free, the minimization condition $\partial_{a}R$=0 results in
\begin{eqnarray}
d_{abc}\,z^{b}z^{c}(q_{0}+3\,d_{def}\,p^{d}z^{e}z^{f})-4\,d_{abc}\,p^{b}z^{c}(d_{def}\,z^{d} z^{e}z^{f})=0\ .
\end{eqnarray}
Making the ansatz $z^{a}=ip^{a}t$ for the attractor values of the moduli immediately solves the above equation
for
\begin{eqnarray}
t=\pm\sqrt{-\frac{q_{0}}{D}}\ ,
\end{eqnarray}
where $D=d_{abc}\,p^{a}p^{b}p^{c}$. This result was first found in \cite{Tripathy:2005qp} by minimizing $V_{BH}$,
and in \cite{Kallosh:2006bt} from the non-supersymmetric version of the attractor equation. In the case with an
arbitrary number of vector multiplets, \cite{Bates:2003vx} offers a prescription for  constructing the full flow
solutions given the attractor values by replacing the charges with corresponding harmonic functions. These flow
equations were explicitly shown to satisfy the first order equation. Hence the same prescription readily applies
for the construction of the non-SUSY solutions. The resulting expressions will satisfy the non-SUSY first order
equations, owing to their formal similarity to the first order SUSY ones. Because an odd number of charges in the
SUSY attractor need to be flipped to get to non-SUSY, we end up with the same solution in terms of harmonics as in
the SUSY case but with an odd number of harmonics having flipped their sign. This is a powerful solution
generating technique for non-SUSY solutions and enables us to obtain them in any scenario where the SUSY solution
is known.

\bibliographystyle{utphys}
\bibliography{nonsusyatt1}

\providecommand{\href}[2]{#2}\begingroup\raggedright\begin{thebibliography}{10}

\bibitem{Dabholkar:2004yr}
A.~Dabholkar, ``{Exact counting of black hole microstates},''
  \href{http://dx.doi.org/10.1103/PhysRevLett.94.241301}{{\em Phys. Rev. Lett.}
  {\bf 94} (2005)  241301},
\href{http://arxiv.org/abs/hep-th/0409148}{{\tt arXiv:hep-th/0409148}}.
%%CITATION = HEP-TH/0409148;%%.

\bibitem{Dabholkar:2005by}
A.~Dabholkar, F.~Denef, G.~W. Moore, and B.~Pioline, ``{Exact and Asymptotic
  Degeneracies of Small Black Holes},'' {\em JHEP} {\bf 08} (2005)  021,
\href{http://arxiv.org/abs/hep-th/0502157}{{\tt arXiv:hep-th/0502157}}.
%%CITATION = HEP-TH/0502157;%%.

\bibitem{Dabholkar:2005dt}
A.~Dabholkar, F.~Denef, G.~W. Moore, and B.~Pioline, ``{Precision counting of
  small black holes},'' {\em JHEP} {\bf 10} (2005)  096,
\href{http://arxiv.org/abs/hep-th/0507014}{{\tt arXiv:hep-th/0507014}}.
%%CITATION = HEP-TH/0507014;%%.

\bibitem{Dabholkar:2008zy}
A.~Dabholkar, J.~Gomes, and S.~Murthy, ``{Counting all dyons in N =4 string
  theory},''
\href{http://arxiv.org/abs/0803.2692}{{\tt arXiv:0803.2692 [hep-th]}}.
%%CITATION = 0803.2692;%%.

\bibitem{Dabholkar:2009dq}
A.~Dabholkar, M.~Guica, S.~Murthy, and S.~Nampuri, ``{No entropy enigmas for
  N=4 dyons},'' \href{http://dx.doi.org/10.1007/JHEP06(2010)007}{{\em JHEP}
  {\bf 06} (2010)  007},
\href{http://arxiv.org/abs/0903.2481}{{\tt arXiv:0903.2481 [hep-th]}}.
%%CITATION = 0903.2481;%%.

\bibitem{Dijkgraaf:1996it}
R.~Dijkgraaf, E.~P. Verlinde, and H.~L. Verlinde, ``{Counting Dyons in N=4
  String Theory},'' \href{http://dx.doi.org/10.1016/S0550-3213(96)00640-2}{{\em
  Nucl. Phys.} {\bf B484} (1997)  543--561},
\href{http://arxiv.org/abs/hep-th/9607026}{{\tt arXiv:hep-th/9607026}}.
%%CITATION = HEP-TH/9607026;%%.

\bibitem{Mandal:2010cj}
I.~Mandal and A.~Sen, ``{Black Hole Microstate Counting and its Macroscopic
  Counterpart},''
\href{http://arxiv.org/abs/1008.3801}{{\tt arXiv:1008.3801 [hep-th]}}.
%%CITATION = 1008.3801;%%.

\bibitem{Pioline:2005vi}
B.~Pioline, ``{BPS black hole degeneracies and minimal automorphic
  representations},'' {\em JHEP} {\bf 08} (2005)  071,
\href{http://arxiv.org/abs/hep-th/0506228}{{\tt arXiv:hep-th/0506228}}.
%%CITATION = HEP-TH/0506228;%%.

\bibitem{Sen:2005ch}
A.~Sen, ``{Black holes and the spectrum of half-BPS states in N = 4
  supersymmetric string theory},'' {\em Adv. Theor. Math. Phys.} {\bf 9} (2005)
   527--558,
\href{http://arxiv.org/abs/hep-th/0504005}{{\tt arXiv:hep-th/0504005}}.
%%CITATION = HEP-TH/0504005;%%.

\bibitem{Sen:2008ta}
A.~Sen, ``{N=8 Dyon Partition Function and Walls of Marginal Stability},''
  \href{http://dx.doi.org/10.1088/1126-6708/2008/07/118}{{\em JHEP} {\bf 07}
  (2008)  118},
\href{http://arxiv.org/abs/0803.1014}{{\tt arXiv:0803.1014 [hep-th]}}.
%%CITATION = 0803.1014;%%.

\bibitem{Shih:2005uc}
D.~Shih, A.~Strominger, and X.~Yin, ``{Recounting dyons in N = 4 string
  theory},'' {\em JHEP} {\bf 10} (2006)  087,
\href{http://arxiv.org/abs/hep-th/0505094}{{\tt arXiv:hep-th/0505094}}.
%%CITATION = HEP-TH/0505094;%%.

\bibitem{Strominger:1990pd}
A.~Strominger, ``{SPECIAL GEOMETRY},''
\href{http://dx.doi.org/10.1007/BF02096559}{{\em Commun. Math. Phys.} {\bf 133}
  (1990)  163--180}.
%%CITATION = CMPHA,133,163;%%.

\bibitem{Dabholkar:2008tm}
A.~Dabholkar, K.~Narayan, and S.~Nampuri, ``{Degeneracy of Decadent Dyons},''
  \href{http://dx.doi.org/10.1088/1126-6708/2008/03/026}{{\em JHEP} {\bf 03}
  (2008)  026},
\href{http://arxiv.org/abs/0802.0761}{{\tt arXiv:0802.0761 [hep-th]}}.
%%CITATION = 0802.0761;%%.

\bibitem{Dabholkar:2006xa}
A.~Dabholkar and S.~Nampuri, ``{Spectrum of Dyons and Black Holes in CHL
  orbifolds using Borcherds Lift},''
  \href{http://dx.doi.org/10.1088/1126-6708/2007/11/077}{{\em JHEP} {\bf 11}
  (2007)  077},
\href{http://arxiv.org/abs/hep-th/0603066}{{\tt arXiv:hep-th/0603066}}.
%%CITATION = HEP-TH/0603066;%%.

\bibitem{Dabholkar:2007vk}
A.~Dabholkar, D.~Gaiotto, and S.~Nampuri, ``{Comments on the spectrum of CHL
  dyons},'' \href{http://dx.doi.org/10.1088/1126-6708/2008/01/023}{{\em JHEP}
  {\bf 01} (2008)  023},
\href{http://arxiv.org/abs/hep-th/0702150}{{\tt arXiv:hep-th/0702150}}.
%%CITATION = HEP-TH/0702150;%%.

\bibitem{Guica:2008mu}
M.~Guica, T.~Hartman, W.~Song, and A.~Strominger, ``{The Kerr/CFT
  Correspondence},'' \href{http://dx.doi.org/10.1103/PhysRevD.80.124008}{{\em
  Phys. Rev.} {\bf D80} (2009)  124008},
\href{http://arxiv.org/abs/0809.4266}{{\tt arXiv:0809.4266 [hep-th]}}.
%%CITATION = 0809.4266;%%.

\bibitem{Jejjala:2009if}
V.~Jejjala and S.~Nampuri, ``{Cardy and Kerr},''
  \href{http://dx.doi.org/10.1007/JHEP02(2010)088}{{\em JHEP} {\bf 02} (2010)
  088},
\href{http://arxiv.org/abs/0909.1110}{{\tt arXiv:0909.1110 [hep-th]}}.
%%CITATION = 0909.1110;%%.

\bibitem{Nampuri:2007gw}
S.~Nampuri, P.~K. Tripathy, and S.~P. Trivedi, ``{Duality Symmetry and the
  Cardy Limit},'' \href{http://dx.doi.org/10.1088/1126-6708/2008/07/072}{{\em
  JHEP} {\bf 07} (2008)  072},
\href{http://arxiv.org/abs/0711.4671}{{\tt arXiv:0711.4671 [hep-th]}}.
%%CITATION = 0711.4671;%%.

\bibitem{Ferrara:1995ih}
S.~Ferrara, R.~Kallosh, and A.~Strominger, ``{N=2 extremal black holes},''
  \href{http://dx.doi.org/10.1103/PhysRevD.52.R5412}{{\em Phys. Rev.} {\bf D52}
  (1995)  5412--5416},
\href{http://arxiv.org/abs/hep-th/9508072}{{\tt arXiv:hep-th/9508072}}.
%%CITATION = HEP-TH/9508072;%%.

\bibitem{Ferrara:1996dd}
S.~Ferrara and R.~Kallosh, ``{Supersymmetry and Attractors},''
  \href{http://dx.doi.org/10.1103/PhysRevD.54.1514}{{\em Phys. Rev.} {\bf D54}
  (1996)  1514--1524},
\href{http://arxiv.org/abs/hep-th/9602136}{{\tt arXiv:hep-th/9602136}}.
%%CITATION = HEP-TH/9602136;%%.

\bibitem{Ferrara:1996um}
S.~Ferrara and R.~Kallosh, ``{Universality of Supersymmetric Attractors},''
  \href{http://dx.doi.org/10.1103/PhysRevD.54.1525}{{\em Phys. Rev.} {\bf D54}
  (1996)  1525--1534},
\href{http://arxiv.org/abs/hep-th/9603090}{{\tt arXiv:hep-th/9603090}}.
%%CITATION = HEP-TH/9603090;%%.

\bibitem{Ferrara:1997tw}
S.~Ferrara, G.~W. Gibbons, and R.~Kallosh, ``{Black holes and critical points
  in moduli space},''
  \href{http://dx.doi.org/10.1016/S0550-3213(97)00324-6}{{\em Nucl. Phys.} {\bf
  B500} (1997)  75--93},
\href{http://arxiv.org/abs/hep-th/9702103}{{\tt arXiv:hep-th/9702103}}.
%%CITATION = HEP-TH/9702103;%%.

\bibitem{Goldstein:2005hq}
K.~Goldstein, N.~Iizuka, R.~P. Jena, and S.~P. Trivedi, ``{Non-supersymmetric
  attractors},'' \href{http://dx.doi.org/10.1103/PhysRevD.72.124021}{{\em Phys.
  Rev.} {\bf D72} (2005)  124021},
\href{http://arxiv.org/abs/hep-th/0507096}{{\tt arXiv:hep-th/0507096}}.
%%CITATION = HEP-TH/0507096;%%.

\bibitem{Tripathy:2005qp}
P.~K. Tripathy and S.~P. Trivedi, ``{Non-Supersymmetric Attractors in String
  Theory},'' {\em JHEP} {\bf 03} (2006)  022,
\href{http://arxiv.org/abs/hep-th/0511117}{{\tt arXiv:hep-th/0511117}}.
%%CITATION = HEP-TH/0511117;%%.

\bibitem{Kallosh:2005ax}
R.~Kallosh, ``{New Attractors},'' {\em JHEP} {\bf 12} (2005)  022,
\href{http://arxiv.org/abs/hep-th/0510024}{{\tt arXiv:hep-th/0510024}}.
%%CITATION = HEP-TH/0510024;%%.

\bibitem{Kallosh:2006bt}
R.~Kallosh, N.~Sivanandam, and M.~Soroush, ``{The non-BPS black hole attractor
  equation},'' {\em JHEP} {\bf 03} (2006)  060,
\href{http://arxiv.org/abs/hep-th/0602005}{{\tt arXiv:hep-th/0602005}}.
%%CITATION = HEP-TH/0602005;%%.

\bibitem{Kallosh:2006bx}
R.~Kallosh, ``{From BPS to non-BPS black holes canonically},''
\href{http://arxiv.org/abs/hep-th/0603003}{{\tt arXiv:hep-th/0603003}}.
%%CITATION = HEP-TH/0603003;%%.

\bibitem{Sen:2007qy}
A.~Sen, ``{Black Hole Entropy Function, Attractors and Precision Counting of
  Microstates},'' \href{http://dx.doi.org/10.1007/s10714-008-0626-4}{{\em Gen.
  Rel. Grav.} {\bf 40} (2008)  2249--2431},
\href{http://arxiv.org/abs/0708.1270}{{\tt arXiv:0708.1270 [hep-th]}}.
%%CITATION = 0708.1270;%%.

\bibitem{Goldstein:2009cv}
K.~Goldstein, S.~Kachru, S.~Prakash, and S.~P. Trivedi, ``{Holography of
  Charged Dilaton Black Holes},''
  \href{http://dx.doi.org/10.1007/JHEP08(2010)078}{{\em JHEP} {\bf 08} (2010)
  078},
\href{http://arxiv.org/abs/0911.3586}{{\tt arXiv:0911.3586 [hep-th]}}.
%%CITATION = 0911.3586;%%.

\bibitem{Goldstein:2010aw}
K.~Goldstein {\em et al.}, ``{Holography of Dyonic Dilaton Black Branes},''
\href{http://arxiv.org/abs/1007.2490}{{\tt arXiv:1007.2490 [hep-th]}}.
%%CITATION = 1007.2490;%%.

\bibitem{Denef:2000nb}
F.~Denef, ``{Supergravity flows and D-brane stability},'' {\em JHEP} {\bf 08}
  (2000)  050,
\href{http://arxiv.org/abs/hep-th/0005049}{{\tt arXiv:hep-th/0005049}}.
%%CITATION = HEP-TH/0005049;%%.

\bibitem{Moore:1998pn}
G.~W. Moore, ``{Arithmetic and attractors},''
\href{http://arxiv.org/abs/hep-th/9807087}{{\tt arXiv:hep-th/9807087}}.
%%CITATION = HEP-TH/9807087;%%.

\bibitem{Moore:1998zu}
G.~W. Moore, ``{Attractors and arithmetic},''
\href{http://arxiv.org/abs/hep-th/9807056}{{\tt arXiv:hep-th/9807056}}.
%%CITATION = HEP-TH/9807056;%%.

\bibitem{Bates:2003vx}
B.~Bates and F.~Denef, ``{Exact solutions for supersymmetric stationary black
  hole composites},''
\href{http://arxiv.org/abs/hep-th/0304094}{{\tt arXiv:hep-th/0304094}}.
%%CITATION = HEP-TH/0304094;%%.

\bibitem{Behrndt:1997ny}
K.~Behrndt, D.~Lust, and W.~A. Sabra, ``{Stationary solutions of N = 2
  supergravity},'' \href{http://dx.doi.org/10.1016/S0550-3213(97)00633-0}{{\em
  Nucl. Phys.} {\bf B510} (1998)  264--288},
\href{http://arxiv.org/abs/hep-th/9705169}{{\tt arXiv:hep-th/9705169}}.
%%CITATION = HEP-TH/9705169;%%.

\bibitem{Ceresole:2007wx}
A.~Ceresole and G.~Dall'Agata, ``{Flow Equations for Non-BPS Extremal Black
  Holes},'' {\em JHEP} {\bf 03} (2007)  110,
\href{http://arxiv.org/abs/hep-th/0702088}{{\tt arXiv:hep-th/0702088}}.
%%CITATION = HEP-TH/0702088;%%.

\bibitem{Bellucci:2008sv}
S.~Bellucci, S.~Ferrara, A.~Marrani, and A.~Yeranyan, ``{stu Black Holes
  Unveiled},''
\href{http://arxiv.org/abs/0807.3503}{{\tt arXiv:0807.3503 [hep-th]}}.
%%CITATION = 0807.3503;%%.

\bibitem{Ceresole:2009vp}
A.~Ceresole, G.~Dall'Agata, S.~Ferrara, and A.~Yeranyan, ``{Universality of the
  superpotential for d = 4 extremal black holes},''
\href{http://arxiv.org/abs/0910.2697}{{\tt arXiv:0910.2697 [hep-th]}}.
%%CITATION = 0910.2697;%%.

\bibitem{Andrianopoli:2010bj}
L.~Andrianopoli, R.~D'Auria, S.~Ferrara, and M.~Trigiante, ``{Fake
  Superpotential for Large and Small Extremal Black Holes},''
  \href{http://dx.doi.org/10.1007/JHEP08(2010)126}{{\em JHEP} {\bf 08} (2010)
  126},
\href{http://arxiv.org/abs/1002.4340}{{\tt arXiv:1002.4340 [hep-th]}}.
%%CITATION = 1002.4340;%%.

\bibitem{Andrianopoli:2007gt}
L.~Andrianopoli, R.~D'Auria, E.~Orazi, and M.~Trigiante, ``{First Order
  Description of Black Holes in Moduli Space},''
  \href{http://dx.doi.org/10.1088/1126-6708/2007/11/032}{{\em JHEP} {\bf 11}
  (2007)  032},
\href{http://arxiv.org/abs/0706.0712}{{\tt arXiv:0706.0712 [hep-th]}}.
%%CITATION = 0706.0712;%%.

\bibitem{Cardoso:2007ky}
G.~Lopes~Cardoso, A.~Ceresole, G.~Dall'Agata, J.~M. Oberreuter, and J.~Perz,
  ``{First-order flow equations for extremal black holes in very special
  geometry},'' \href{http://dx.doi.org/10.1088/1126-6708/2007/10/063}{{\em
  JHEP} {\bf 10} (2007)  063},
\href{http://arxiv.org/abs/0706.3373}{{\tt arXiv:0706.3373 [hep-th]}}.
%%CITATION = 0706.3373;%%.

\bibitem{Galli:2009bj}
P.~Galli and J.~Perz, ``{Non-supersymmetric extremal multicenter black holes
  with superpotentials},'' {\em JHEP} {\bf 02} (2010)  102,
\href{http://arxiv.org/abs/0909.5185}{{\tt arXiv:0909.5185 [hep-th]}}.
%%CITATION = 0909.5185;%%.

\bibitem{Brunner:2010xm}
I.~Brunner and D.~Roggenkamp, ``{Attractor Flows from Defect Lines},''
\href{http://arxiv.org/abs/1002.2614}{{\tt arXiv:1002.2614 [hep-th]}}.
%%CITATION = 1002.2614;%%.

\bibitem{Ceresole:2010hq}
A.~Ceresole and S.~Ferrara, ``{Black Holes and Attractors in Supergravity},''
\href{http://arxiv.org/abs/1009.4175}{{\tt arXiv:1009.4175 [hep-th]}}.
%%CITATION = 1009.4175;%%.

\bibitem{Ceresole:2009id}
A.~Ceresole, S.~Ferrara, and A.~Gnecchi, ``{5d/4d U-dualities and N=8 black
  holes},'' \href{http://dx.doi.org/10.1103/PhysRevD.80.125033}{{\em Phys.
  Rev.} {\bf D80} (2009)  125033},
\href{http://arxiv.org/abs/0908.1069}{{\tt arXiv:0908.1069 [hep-th]}}.
%%CITATION = 0908.1069;%%.

\bibitem{Andrianopoli:2008ea}
L.~Andrianopoli, R.~D'Auria, S.~Ferrara, P.~A. Grassi, and M.~Trigiante,
  ``{Exceptional N=6 and N=2 AdS4 Supergravity, and Zero- Center Modules},''
  \href{http://dx.doi.org/10.1088/1126-6708/2009/04/074}{{\em JHEP} {\bf 04}
  (2009)  074},
\href{http://arxiv.org/abs/0810.1214}{{\tt arXiv:0810.1214 [hep-th]}}.
%%CITATION = 0810.1214;%%.

\bibitem{Ferrara:2010ru}
S.~Ferrara and A.~Marrani, ``{Matrix Norms, BPS Bounds and Marginal Stability
  in N=8 Supergravity},''
\href{http://arxiv.org/abs/1009.3251}{{\tt arXiv:1009.3251 [hep-th]}}.
%%CITATION = 1009.3251;%%.

\bibitem{Kallosh:2006ib}
R.~Kallosh, N.~Sivanandam, and M.~Soroush, ``{Exact attractive non-BPS STU
  black holes},'' \href{http://dx.doi.org/10.1103/PhysRevD.74.065008}{{\em
  Phys. Rev.} {\bf D74} (2006)  065008},
\href{http://arxiv.org/abs/hep-th/0606263}{{\tt arXiv:hep-th/0606263}}.
%%CITATION = HEP-TH/0606263;%%.

\bibitem{Kallosh:2006zs}
R.~Kallosh and A.~D. Linde, ``{Strings, black holes, and quantum
  information},'' \href{http://dx.doi.org/10.1103/PhysRevD.73.104033}{{\em
  Phys. Rev.} {\bf D73} (2006)  104033},
\href{http://arxiv.org/abs/hep-th/0602061}{{\tt arXiv:hep-th/0602061}}.
%%CITATION = HEP-TH/0602061;%%.

\bibitem{Gimon:2007mh}
E.~G. Gimon, F.~Larsen, and J.~Simon, ``{Black Holes in Supergravity: the
  non-BPS Branch},''
  \href{http://dx.doi.org/10.1088/1126-6708/2008/01/040}{{\em JHEP} {\bf 01}
  (2008)  040},
\href{http://arxiv.org/abs/0710.4967}{{\tt arXiv:0710.4967 [hep-th]}}.
%%CITATION = 0710.4967;%%.

\bibitem{Nampuri:2007gv}
S.~Nampuri, P.~K. Tripathy, and S.~P. Trivedi, ``{On The Stability of
  Non-Supersymmetric Attractors in String Theory},''
  \href{http://dx.doi.org/10.1088/1126-6708/2007/08/054}{{\em JHEP} {\bf 08}
  (2007)  054},
\href{http://arxiv.org/abs/0705.4554}{{\tt arXiv:0705.4554 [hep-th]}}.
%%CITATION = 0705.4554;%%.

\bibitem{Andriyash:2010qv}
E.~Andriyash, F.~Denef, D.~L. Jafferis, and G.~W. Moore, ``{Wall-crossing from
  supersymmetric galaxies},''
\href{http://arxiv.org/abs/1008.0030}{{\tt arXiv:1008.0030 [hep-th]}}.
%%CITATION = 1008.0030;%%.

\bibitem{Andriyash:2010yf}
E.~Andriyash, F.~Denef, D.~L. Jafferis, and G.~W. Moore, ``{Bound state
  transformation walls},''
\href{http://arxiv.org/abs/1008.3555}{{\tt arXiv:1008.3555 [hep-th]}}.
%%CITATION = 1008.3555;%%.

\bibitem{Bena:2009qv}
I.~Bena, S.~Giusto, C.~Ruef, and N.~P. Warner, ``{A (Running) Bolt for New
  Reasons},'' \href{http://dx.doi.org/10.1088/1126-6708/2009/11/089}{{\em JHEP}
  {\bf 0911} (2009)  089}, \href{http://arxiv.org/abs/0909.2559}{{\tt
  arXiv:0909.2559 [hep-th]}}.

\bibitem{Bena:2009fi}
I.~Bena, S.~Giusto, C.~Ruef, and N.~P. Warner, ``{Supergravity Solutions from
  Floating Branes},'' \href{http://dx.doi.org/10.1007/JHEP03(2010)047}{{\em
  JHEP} {\bf 1003} (2010)  047}, \href{http://arxiv.org/abs/0910.1860}{{\tt
  arXiv:0910.1860 [hep-th]}}.

\bibitem{Bena:2009en}
I.~Bena, S.~Giusto, C.~Ruef, and N.~P. Warner, ``{Multi-Center non-BPS Black
  Holes: the Solution},''
  \href{http://dx.doi.org/10.1088/1126-6708/2009/11/032}{{\em JHEP} {\bf 0911}
  (2009)  032}, \href{http://arxiv.org/abs/0908.2121}{{\tt arXiv:0908.2121
  [hep-th]}}.

\bibitem{Bobev:2009kn}
N.~Bobev and C.~Ruef, ``{The Nuts and Bolts of Einstein-Maxwell Solutions},''
  \href{http://dx.doi.org/10.1007/JHEP01(2010)124}{{\em JHEP} {\bf 1001} (2010)
   124}, \href{http://arxiv.org/abs/0912.0010}{{\tt arXiv:0912.0010 [hep-th]}}.

\bibitem{Ceresole:2007rq}
A.~Ceresole, S.~Ferrara, and A.~Marrani, ``{4d/5d Correspondence for the Black
  Hole Potential and its Critical Points},''
  \href{http://dx.doi.org/10.1088/0264-9381/24/22/023}{{\em Class. Quant.
  Grav.} {\bf 24} (2007)  5651--5666},
\href{http://arxiv.org/abs/0707.0964}{{\tt arXiv:0707.0964 [hep-th]}}.
%%CITATION = 0707.0964;%%.

\bibitem{Bachas:2007td}
C.~Bachas and I.~Brunner, ``{Fusion of conformal interfaces},''
  \href{http://dx.doi.org/10.1088/1126-6708/2008/02/085}{{\em JHEP} {\bf 02}
  (2008)  085},
\href{http://arxiv.org/abs/0712.0076}{{\tt arXiv:0712.0076 [hep-th]}}.
%%CITATION = 0712.0076;%%.

\bibitem{Brunner:2008fa}
I.~Brunner, H.~Jockers, and D.~Roggenkamp, ``{Defects and D-Brane
  Monodromies},'' {\em Adv. Theor. Math. Phys.} {\bf 13} (2009)  1077--1135,
\href{http://arxiv.org/abs/0806.4734}{{\tt arXiv:0806.4734 [hep-th]}}.
%%CITATION = 0806.4734;%%.

\bibitem{Brunner:2007ur}
I.~Brunner and D.~Roggenkamp, ``{Defects and Bulk Perturbations of Boundary
  Landau-Ginzburg Orbifolds},''
  \href{http://dx.doi.org/10.1088/1126-6708/2008/04/001}{{\em JHEP} {\bf 04}
  (2008)  001},
\href{http://arxiv.org/abs/0712.0188}{{\tt arXiv:0712.0188 [hep-th]}}.
%%CITATION = 0712.0188;%%.

\bibitem{Bachas:2001vj}
C.~Bachas, J.~de~Boer, R.~Dijkgraaf, and H.~Ooguri, ``{Permeable conformal
  walls and holography},'' {\em JHEP} {\bf 06} (2002)  027,
\href{http://arxiv.org/abs/hep-th/0111210}{{\tt arXiv:hep-th/0111210}}.
%%CITATION = HEP-TH/0111210;%%.

\end{thebibliography}\endgroup

\end{document}